\documentclass{aa} 

\usepackage{graphicx}
 \usepackage[varg]{txfonts}
\usepackage{natbib}
\usepackage{hyperref}
\hypersetup{colorlinks=true,allcolors=blue,breaklinks=true}
\usepackage{xcolor}
\usepackage{amssymb,amsmath}
\usepackage[varg]{txfonts}
\usepackage{graphicx}
\usepackage{lscape}
\usepackage{array}
\usepackage{xspace}
\usepackage{comment}
\usepackage[frozencache,cachedir=.]{minted}
\newcommand{\chandra}{\textit{Chandra}\xspace}
\newcommand{\athena}{\textit{Athena}\xspace}
\newcommand{\xrism}{\textit{XRISM}\xspace}

\newcommand{\flux}{ph cm$^{-2}$ s$^{-1}$}
\newcommand{\ergs}{erg s$^{-1}$}
\newcommand{\vel}{km s$^{-1}$}

\begin{document}

\title{Looking through the photoionisation wake: Vela X-1 at $\varphi_\mathrm{orb} \approx0.75$ with \textsl{Chandra}/HETG}

   \author{
R.~Amato\inst{1,2,3}\and
V.~Grinberg\inst{3} \and 
N.~Hell\inst{4} \and
S.~Bianchi\inst{5} \and
C.~Pinto\inst{2} \and
A.~D'Aì\inst{2} \and
M.~Del Santo\inst{2} \and \\
T.~Mineo\inst{2} \and
A.~Santangelo\inst{3}
}

\institute{Dipartimento di Fisica e Chimica - Emilio Segrè, Università degli Studi di Palermo, via Archirafi, 36, 90123 Palermo, Italy
    \and INAF-IASF Palermo, via Ugo la Malfa, 153, 90146 Palermo, Italy
    \and Institute for Astronomy and Astrophysics (IAAT), University of T\"ubingen, Sand 1, 72076 T\"ubingen, Germany 
    \and Lawrence Livermore National Laboratory, 7000 East Ave, Livermore, CA 94550, USA
    \and Dipartimento di Matematica e Fisica, Università degli Studi Roma Tre, Largo S. Leonardo Murialdo, 1, 00146 Roma, Italy
}

\date{ -- / --}

\abstract{The Supergiant X-ray binary Vela X-1 represents one of the best astrophysical sources to investigate the wind environment of a O/B star irradiated by an accreting neutron star. Previous studies and hydrodynamic simulations of the system revealed a clumpy environment and the presence of two wakes: an accretion wake surrounding the compact object and a photoionisation wake trailing it along the orbit.}{Our goal is to conduct, for the first time, high-resolution spectroscopy on \chandra/HETGS data at the orbital phase $\varphi_\mathrm{orb} \approx 0.75$, when the line of sight is crossing the photoionisation wake. We aim to conduct plasma diagnostics, inferring the structure and the geometry of the wind.}{We perform a blind search employing a Bayesian Block algorithm to find discrete spectral features
and identify them thanks to the most recent laboratory results or through atomic databases. Plasma properties are inferred both with empirical techniques and with photoionisation models within CLOUDY and SPEX.}{We detect and identify five narrow radiative recombination continua (\ion{Mg}{xi-xii}, \ion{Ne}{ix-x}, \ion{O}{viii}) and several emission lines from Fe, S, Si, Mg, Ne, Al, and Na, including four He-like triplets (\ion{S}{xv}, \ion{Si}{xiii}, \ion{Mg}{xi}, and \ion{Ne}{ix}). Photoionisation models well reproduce the overall spectrum, except for the near-neutral fluorescence lines of Fe, S, and Si.}{We conclude that the plasma is mainly photoionised, but more than one component is most likely present, consistent with a multi-phase plasma scenario, where denser and colder clumps of matter are embedded in the hot, photoionised wind of the companion star. Simulations with the future X-ray satellites \athena\ and \xrism\ show that a few hundred seconds of exposure will be sufficient to disentangle the lines of the Fe K$\alpha$ doublet and the He-like \ion{Fe}{XXV}, improving, in general, the determination of the plasma parameters.}


\maketitle
\section{Introduction}
\label{sec:intro}
\begin{figure}
\centering
\includegraphics[width=.35\textwidth]{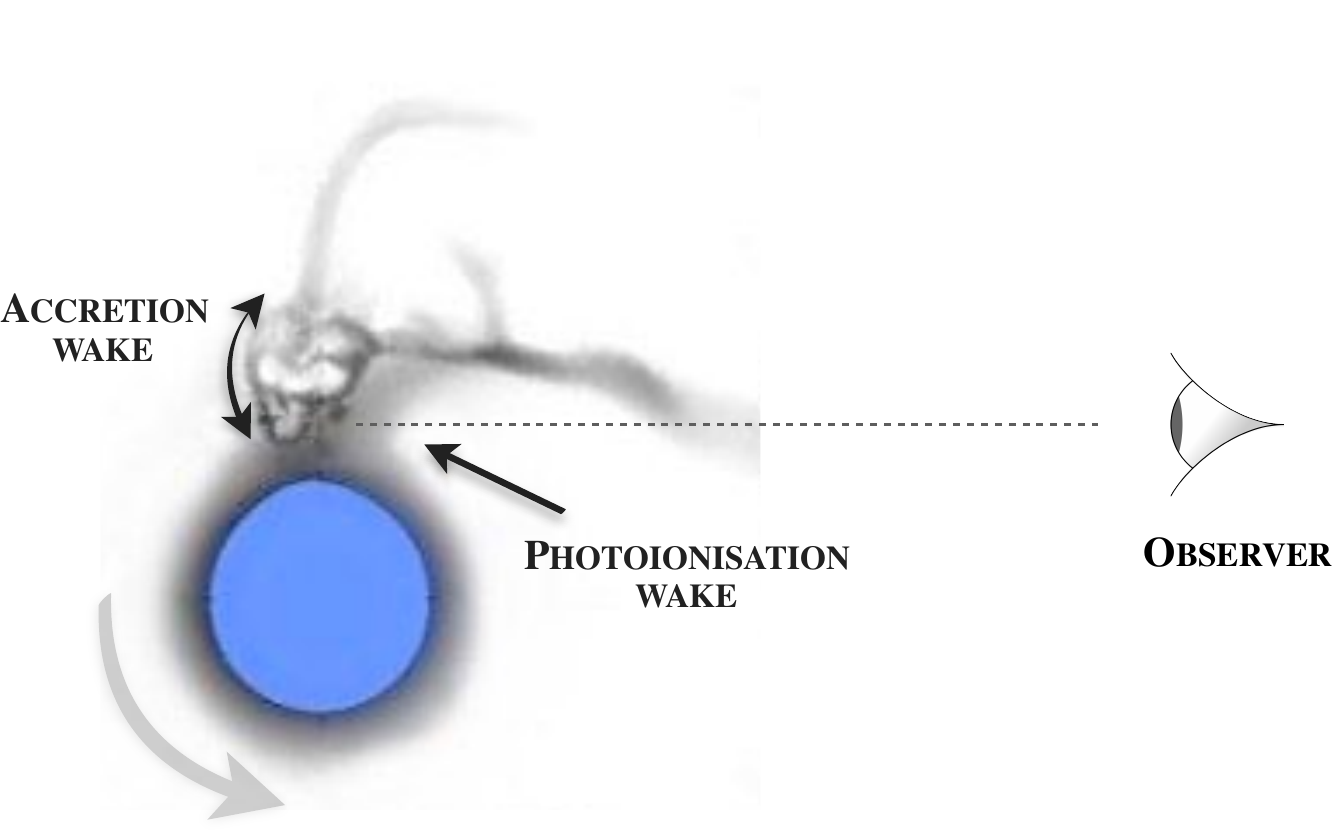}
      \caption{A sketch of Vela X-1 from \citet{Grinberg_2017a} showing the accretion and photoionisation wakes. The blue circle represents the donor star HD 77851, while the pulsar is hidden in the accretion wake. The grey arrow indicates the verse of the rotation of the binary system. At the orbital phase $\varphi_{\text{orb}}\approx0.75$, the observer is looking at the system from the right, so that the line of sight (dashed line) is crossing the photoionisation wake.}
     \label{fig:sketch_velax1}
   \end{figure}

The eclipsing high-mass X-ray binary (HMXB) Vela X-1 (4U~0900-40) consists of a $\sim$283 s period pulsar  \citep{McClintock1976} and a blue supergiant companion star \citep[HD 77851, a B0.5Ia class star,][]{Hiltner1972}. With an X-ray luminosity of $\sim4\times10^{36}$ \ergs\ and a distance of 2 kpc from Earth \citep{Gimenez-Garcia2016}, it is one of the brightest HMXBs in the sky. It is a high inclination system \citep[>73$^\circ$,][]{Joss1984}, with an orbital period of $\sim$8.9 d \citep{Forman1973,Kreykenbohm2008} and an orbital separation of $\sim$53\,$R_\sun$ \citep{Quaintrell2003}. The donor star has a radius of about 30\,$R_\sun$ \citep{Quaintrell2003}, so that the pulsar is constantly embedded in the wind environment of the companion. The geometry of the 
accreting stream of matter onto the compact object is complex, being made up of an accretion wake, a photoionisation wake,
and possibly a tidal stream, 
as both simulations  \citep[e.g.,][]{Blondin1990,Manousakis_2011_PhD} and observations in different wavebands show \citep[e.g.,][]{Kaper_1994a,van_Loon_2001a,Malacaria2016}. A sketch of the binary system with the main features is given in Fig.\,\ref{fig:sketch_velax1}. The line of sight intersects the different elements at different orbital phases, so that the observational data show strong changes in absorption along the whole orbital period \citep{Doroshenko2013}.

X-ray emission from Vela X-1 has already been detected and studied for several different orbital phases with different instruments \citep[e.g.,][]{Haberl1990,Goldstein2004,Watanabe2006,Furst2010,Grinberg_2017a}. High resolution X-ray studies of the system are of special interest, as they allow to draw conclusions on the properties of the complex plasma. High-resolution data from the High-Energy Transmission Grating Spectrometer (HETGS) \citep{Canizares2005} of the \textit{Chandra X-ray Observatory} \citep{Weisskopf2000} of Vela X-1 during eclipse ($\varphi_\mathrm{orb}\approx 0$) were studied by \citet{Schulz2002}, who discovered and identified a variety of emission features, including radiative recombination continua (RRCs) and fluorescent lines, that led to the idea of the coexistence of a hot optically thin photoionised plasma and a colder optically thick one. \citet{Goldstein2004} investigated \chandra/HETGS data of the system at three different orbital phases ($\varphi_\mathrm{orb}\approx0$, $\varphi_\mathrm{orb}\approx0.25$, $\varphi_\mathrm{orb}\approx0.5$), finding that the emission features revealed during the eclipse are obscured at $\varphi_\mathrm{orb}\approx0.25$, but then they appear again at $\varphi_\mathrm{orb}\approx0.5$, when the soft X-ray continuum diminishes. The simultaneous presence of H- and He-like emission lines and fluorescent lines of near-neutral ions can be explained by contributions from different regions: the warm photoionised wind of the companion star and smaller cooler regions, or clumps, of gas.
\citet{Watanabe2006} compared the same \chandra/HETGS data sets to 3D Monte Carlo simulations of X-ray photons propagating through a smooth, undisturbed wind. Based on this assumption, they deducted that highly ionised ions, which give rise to the emission lines, are located mainly in the region between the neutron star (NS) and the companion star, while the fluorescent lines are produced in the extended stellar wind, from reflection of the stellar photosphere, and in the accretion wake. More recent results on the same orbital phase by \citet{Odaka2013} with \textit{Suzaku} and by \citet{MartinezNunez2014} with \textit{XMM-Newton}, respectively, highlighted flux variability and strong changes in absorption over periods of the order of ks. The same variability is found in \chandra/HETGS data at $\varphi_\mathrm{orb}\approx0.25$ from \citet{Grinberg_2017a}, 
who attributed the changes in the overall absorption necessarily to the clumpy nature of the winds from the companion. Moreover, the high energy resolution of \chandra allowed the detection of line emission features from several ionised elements, corroborating the idea of a co-existence of cool and hot gas phases in the system.

Hydrodynamic simulations \citep{Manousakis2015,ElMellah2018,ElMellah2019} suggest the presence of a more complex structure around the neutron star (NS), with a bow shock and eventually the formation of a transient wind-captured accretion disk \citep{Liao2020}. Such features can influence the way clumps accrete onto the compact objects, i.e., reducing the amount of transferred angular momentum or introducing time lags and phase mixing when the clumps are stored in such structures.

In this work we present, for the first time, a high-resolution spectroscopic study of \chandra/HETG archival data of Vela X-1 at orbital phase $\varphi_\mathrm{orb}\approx 0.75$, i.e., when the line of sight is intersecting the photoionisation wake (see Fig.\,\ref{fig:sketch_velax1}). The study of the X-ray spectrum at this specific orbital phase, where the absorption from the wind of the X-rays coming from the NS is high, allows the detection of a large number of lines from different elements in high ionisation states and, thus, the application of plasma diagnostic techniques to characterise the accretion environment. 
The paper is structured as follows: we first look for changes in the hardness of the flux in Section \ref{sec:data_reduction}, finding none; then we proceed with a blind search for spectroscopic absorption/emission features, applying a Bayesian Block algorithm to the unbinned spectrum; we present the identification of all the detected features in Section \ref{sec:high_resolution_spectroscopy}, while in Section \ref{cloudy_spex_simulations} we compare the observational data with two different photoionisation codes; in Section \ref{sec:discussion} we discuss the plasma properties and the geometry of the wind of the companion star; in Section \ref{athena_xrism_simulations} we perform simulations with future X-ray satellites; we present our conclusions in Section \ref{sec:conclusion}.

\section{Data reduction and temporal analysis}
\label{sec:data_reduction}
We analysed the High Energy Grating (HEG) and Medium Energy Grating (MEG) data sets of the \chandra{}/HETG ObsID 14654, taken on 2013-07-30, with ACIS-S, in FAINT mode, for a total exposure time of 45.88 ks. According to the ephemeris of \citet{Kreykenbohm2008}, the data set covers the orbital phase $\varphi_\mathrm{orb}=0.72-0.78$, where $\varphi_\mathrm{orb}=0$ is defined as mid-eclipse.  
Data were reprocessed using CIAO 4.11, with CALDB 4.8.2. We followed the standard \chandra\ data analysis threads, but we chose a narrower sky mask to avoid the overlapping of the extraction region and to improve the flux at the shortest wavelengths. 

\begin{figure}
\centering
\includegraphics[width=\hsize]{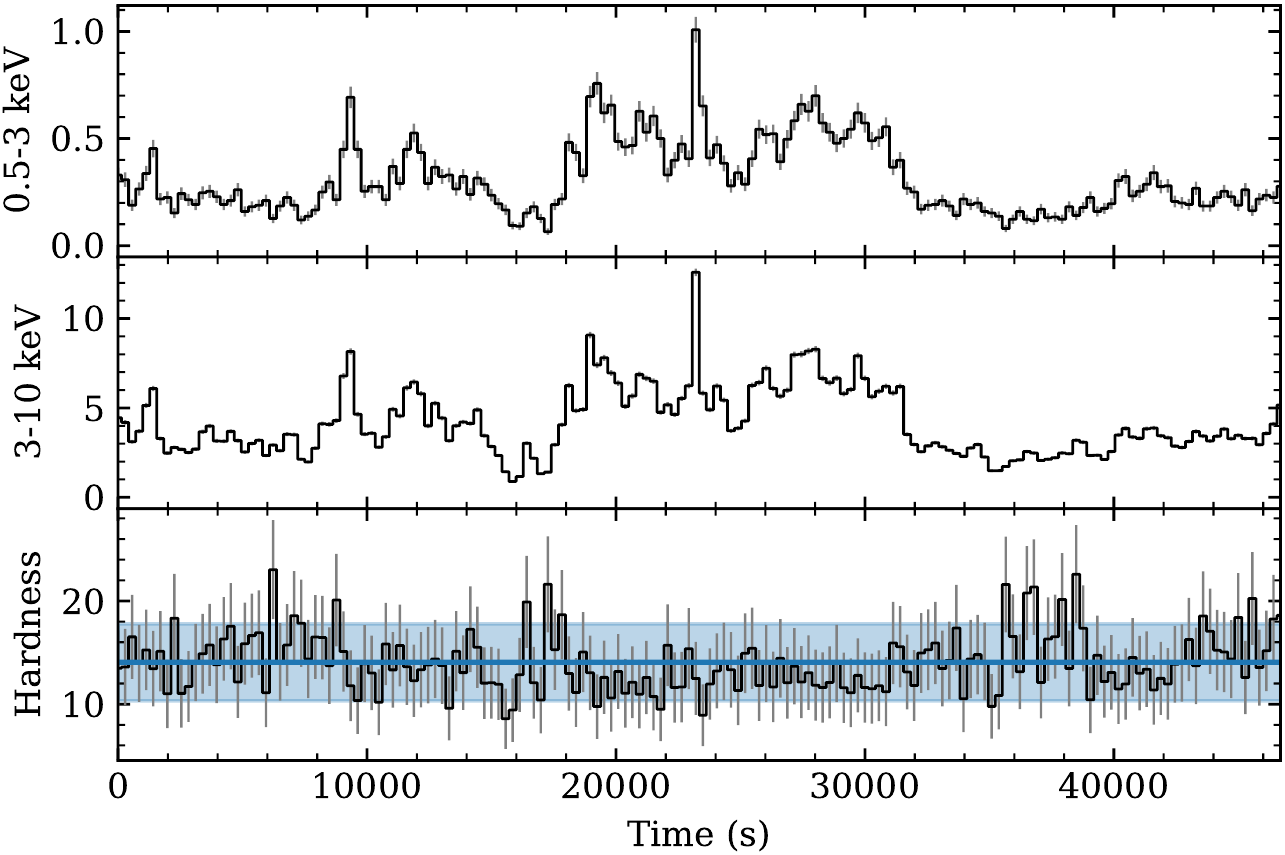}
      \caption{Light curves in units of counts\,s$^{-1}$, in the soft 0.5-3\,keV (\textit{top panel}) and hard 3-10\,keV  (\textit{middle panel}) energy bands, and corresponding hardness ratio ((3--10\,keV)/(0.5--3\,keV), \textit{bottom panel}). The blue horizontal line indicates the mean value of the hardness ratio, with the 1$\sigma$ uncertainty given by the blue area. Data are binned to the spin period of 238 s, error bars at 1$\sigma$.}
         \label{fig:lcurve}
   \end{figure}

\begin{figure}
\centering
\includegraphics[width=\hsize]{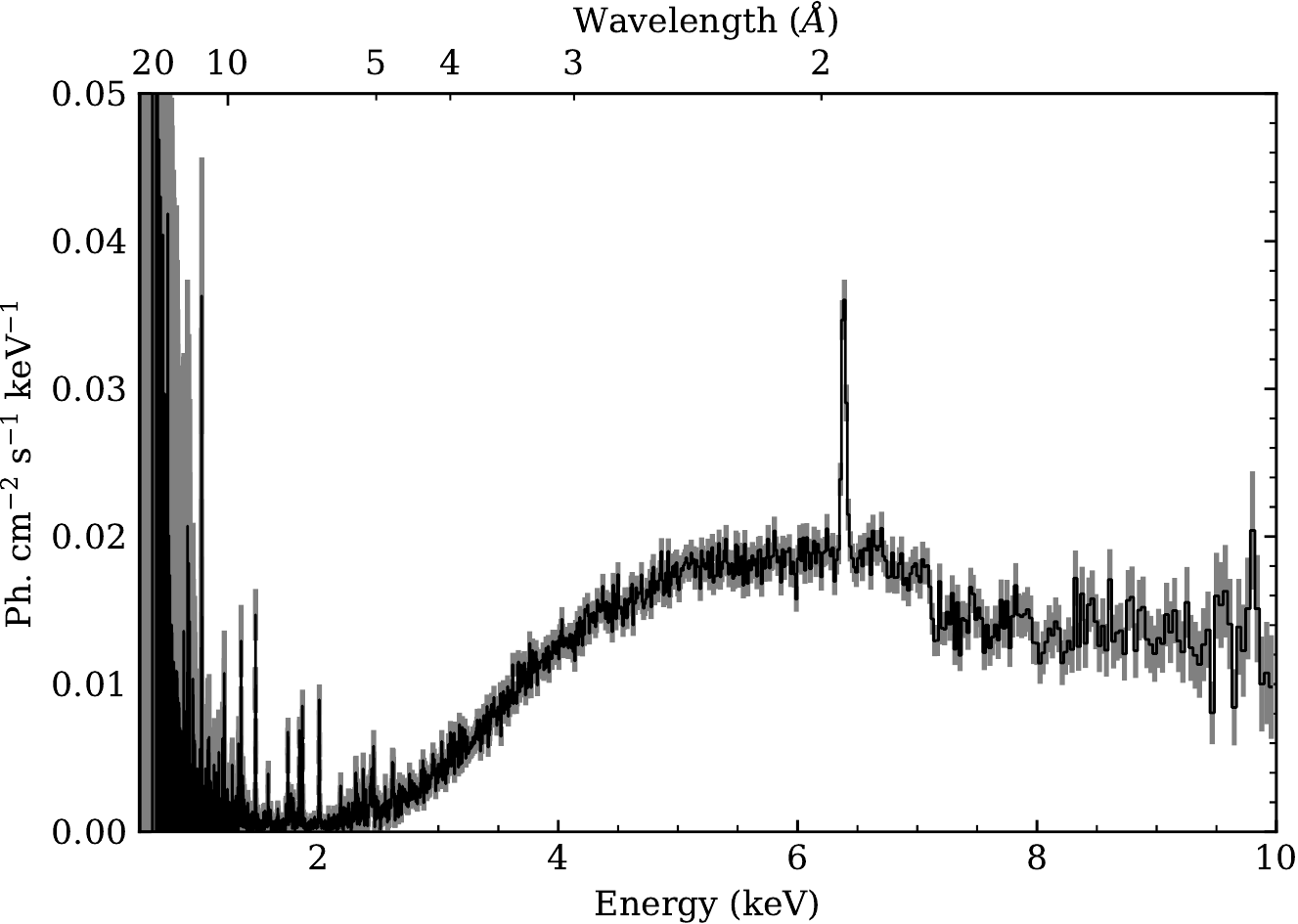}
      \caption{Combined HEG and MEG spectrum of \chandra\ ObsID 14654 in the energy range 0.5-10 keV.}
     \label{fig:spectrum}
   \end{figure}

Following the work of \citet{Grinberg_2017a}, who observed a change in the hardness of the source during phase $\varphi_\mathrm{orb}\approx0.25$, we extracted the light curve in two different energy bands, 0.5--3 keV (soft) and 3--10 keV (hard), and computed the hardness ratio, defined as the ratio between the counts in the hard and soft bands. Fig.\,\ref{fig:lcurve} shows the result, with data binned to the neutron star spin period of 283 s (errors at 1$\sigma$). The hardness ratio values at $\varphi_\mathrm{orb}\approx0.75$ are higher than the ones obtained by \citet{Grinberg_2017a} by at least a factor of ten,  which is not surprising considered the high absorption expected at this orbital phase. Moreover, the hardness ratio is almost flat for the whole observation, in contrast to \citet{Grinberg_2017a}, where a variability of a factor of three was observed.  Hence, we extract only one spectrum, in the full energy range of 0.5--10 keV (Fig.\,\ref{fig:spectrum}). 

\section{High-resolution spectroscopy}
\label{sec:high_resolution_spectroscopy}

We used the Interactive
Spectral Interpretation System (ISIS) 1.6.2-43 \citep{Noble2008,NobleNowak2008} to perform the spectroscopic analysis of the data,  with the ISIS functions (ISISscripts) provided by ECAP/Remeis observatory and MIT\footnote{\url{http://www.sternwarte.uni-erlangen.de/isis/}.}, cross sections from \citet{Verner1996}, and solar abundances from \citet{wilms2000}. We used Cash statistic \citep{Cash1979} with the spectrum binned to the MEG resolution. All uncertainties are given at 90\% confidence level.

We performed a blind search of spectral features, using a Bayesian Block (BB) algorithm \citep{Scargle2013}, as described in \citet{Young2007} and as applied to \chandra/HETGS data by \citet{Grinberg_2017a}. To optimise the line detection algorithm, we divided the whole spectrum into five regions of interest, named after the most significant element detected in each of them, as reported in Table \ref{tab:powerlaw_bestfit}. These spectral regions were analysed one by one. We locally modelled the continuum with a simple power law and then looked for significant deviations in the residuals. Given the narrow wavelength ranges of individual regions, the continuum is always adequately well fitted by a power law. Similar piece-wise approaches have been previously repeatedly used for line searches and modelling \citep[see, e.g.,][]{Yao_2008a, van_den_Ejnden_2019a}.

The BB algorithm determines whether a data point is far from the model above a certain significant threshold, defined by a parameter, $\alpha$, such that each detection has a significance of $p\sim\exp(-2\alpha)$, corresponding to a probability of $P\sim1-\exp(-2\alpha)$ of positive detection. For each new detection, we added to the model one or more Gaussian components for emission/absorption lines and the XSPEC \citep{Xspec1996} functions \mintinline{python}{redge} and \mintinline{python}{edge} for the RRCs and the Fe K-edge (Sect.\,\ref{sec:iron}, \ref{sub:Fe_region}), respectively. After each addition, we fit the data and apply the algorithm once more. We iterate the process until the significance drops to 95\%, corresponding to $\alpha\sim$1.5. All the line detections with their corresponding values of $\alpha$ are listed in Tables \ref{tab:Fe}--\ref{tab:Ne}, while Table \ref{tab:powerlaw_bestfit} shows the best-fit values of the power law parameters for each spectral region and the goodness of the fit. Table \ref{tab:RRC} displays the best-fit values of the RRCs.

In some cases, lines that are too close to be clearly resolved by the algorithm, such as for example He-like triplets, are detected as a single block. In such cases, we use our knowledge of atomic physics to add the proper number of lines to the model. Moreover, to improve the fit, we fixed the distance of known lines, since the BB per se is a blind algorithm, i.e., it does not take into account known line distances. We did this for the H-like Ly$\alpha$ and Ly$\beta$ lines \citep{Erickson1977} and the He-like triplets \citep{Drake1988}, assuming that Doppler
shifts are the same within the same ionic species. 

Whenever a line appeared unresolved, we fixed its width to 0.003~\AA, corresponding to about one third of the MEG resolution (0.023~\AA\ FWHM)\footnote{\url{http://cxc.harvard.edu/cdo/about_chandra}}. Line identification for S and Si ions accounts for the most recent laboratory measurements from \citet{Hell2016}, while for the other elements we use the AtomDB database\footnote{\url{http://www.atomdb.org/index.php}} \citep{Foster2012,Foster2017}.

For every detected He-triplet, we computed the density-sensitive ratio $R=f/i$ and the temperature-sensitive ratio $G=(i+f)/r$, where $f$ represents the intensity of the forbidden line (1s2s $^3$S$_1$--1s$^2$ $^1$S$_0$), $i$ the intensity of the intercombination line (1s2p $^3$P$_1$--1s$^2$ $^1$S$_0$) and $r$ the intensity of the resonant line (1s2p $^1$P$_1$--1s$^2$ $^1$S$_0$) \citep{GabrielJordan1969,PorquetDubau2000}\footnote{\citet{Gabriel1972} refers to the transitions of the lines of the He-like triplets as $w$ for the resonant line, $x$ and $y$ for the two components of the intercombination line, and $z$ for the forbidden line. With this notation, the ratios for plasma diagnostic are expressed as $R=z/(x+y)$ and $G=(z+(x+y))/w$.}. In our case, the intensities of the lines are linked to reproduce $G$ and $R$ as free parameters in the fit. Results are reported in Table~\ref{tab:R_G_ratios}.

In the following subsections, we present in detail the results of the BB procedure for each spectral region of interest.

\begin{table*}
\renewcommand{\arraystretch}{1.3}
\caption{Best-fit values of the power laws used to model the continuum and values of the Cash statistic per degrees of freedom (d.o.f.) for each region of the spectrum.}
\centering
\begin{tabular}{ccccc}
\hline\hline
Region & Wavelength range (\AA) & $\Gamma$ & Norm. (keV s$^{-1}$ cm$^{-2}$) & Cash(d.o.f.)\\
\hline
Fe & 1.6--2.5 & $-0.082^{+0.004}_{-0.008}$ & $0.0158^{+0.0023}_{-0.0018}$ & 1.03(179)\\
S & 4.5--6.0 & $-5.38\pm0.05$ & $\left(1.04^{+0.06}_{-0.16}\right)\times10^{-5}$ & 1.14(279)\\
Si & 6.0--7.4 & $-0.3\pm0.1$ & $\left(3.4\pm0.2\right)\times10^{-4}$ & 1.19(247)\\
Mg & 7.5--10.0 & $2.31^{+0.15}_{-0.16}$ & $\left(1.27\pm0.07\right)\times10^{-3}$ & 1.26(468)\\
Ne & 10.0--14.5 & $0.1\pm0.7$ & $\left(6.9^{+0.7}_{-0.6}\right)\times10^{-4}$ & 0.99(892)\\
\hline\hline
\end{tabular}
\label{tab:powerlaw_bestfit}
\end{table*}

\subsection{Iron region}
\label{sec:iron}

In the Fe region (wavelength range 1.6-2.5\,\AA{}, cf. Table \ref{tab:powerlaw_bestfit}), the BB method found only one strong line, that we identified with Fe K$\alpha$ and one edge, identified with the Fe K-edge. Best-fit values for these features are reported in Table \ref{tab:Fe}. Although the strong Fe K$\alpha$ line implies the presence of a strong Fe K$\beta$ component, our approach did not detect it. We discuss the possible reasons in  Sect.\,\ref{sec:discussion}.

Given the overall strength of the Fe K$\alpha$ line, we attempted an additional fit, letting the line width free. We obtained a best-fit value of $\sigma=\left(3.4^{+0.9}_{-1.1}\right)\times10^{-3}$~\AA, consistent with our previous assumption and with results by \citet{Tzanavaris2018}.

\begin{table*}
\renewcommand{\arraystretch}{1.3}
\caption{Features detected in the Fe region  (1.6--2.5~\AA) with the detection parameter $\alpha$ and the best-fit values. The width of the Fe K$\alpha$ line was fixed to 0.003\AA.}
\centering
\begin{tabular}{cccccc}
\hline\hline
Line & BB  & Ref. wavelength &  Det. wavelength &  Line flux & $\tau$\\
& $\alpha$ & (\AA) & (\AA) & (ph s$^{-1}$ cm$^{-2}$ $\times10^{-4}$) & \\
\hline
Fe K$\alpha$ & 157 & 1.9375\tablefootmark{a} & $1.9388\pm0.0006$ & $9.4\pm0.8$ & -- \\
Fe K edge & 47 & 1.7433\tablefootmark{b} & $1.742\pm0.003$ & -- & $0.31\pm0.03$\\
\hline\hline
\end{tabular}
\tablefoot{
\tablefoottext{a}{\citet{Drake1988}.}
\tablefoottext{b}{\citet{Bearden1967}.}
}
\label{tab:Fe}
\end{table*}

\begin{figure}
\centering
\includegraphics[width=\hsize]{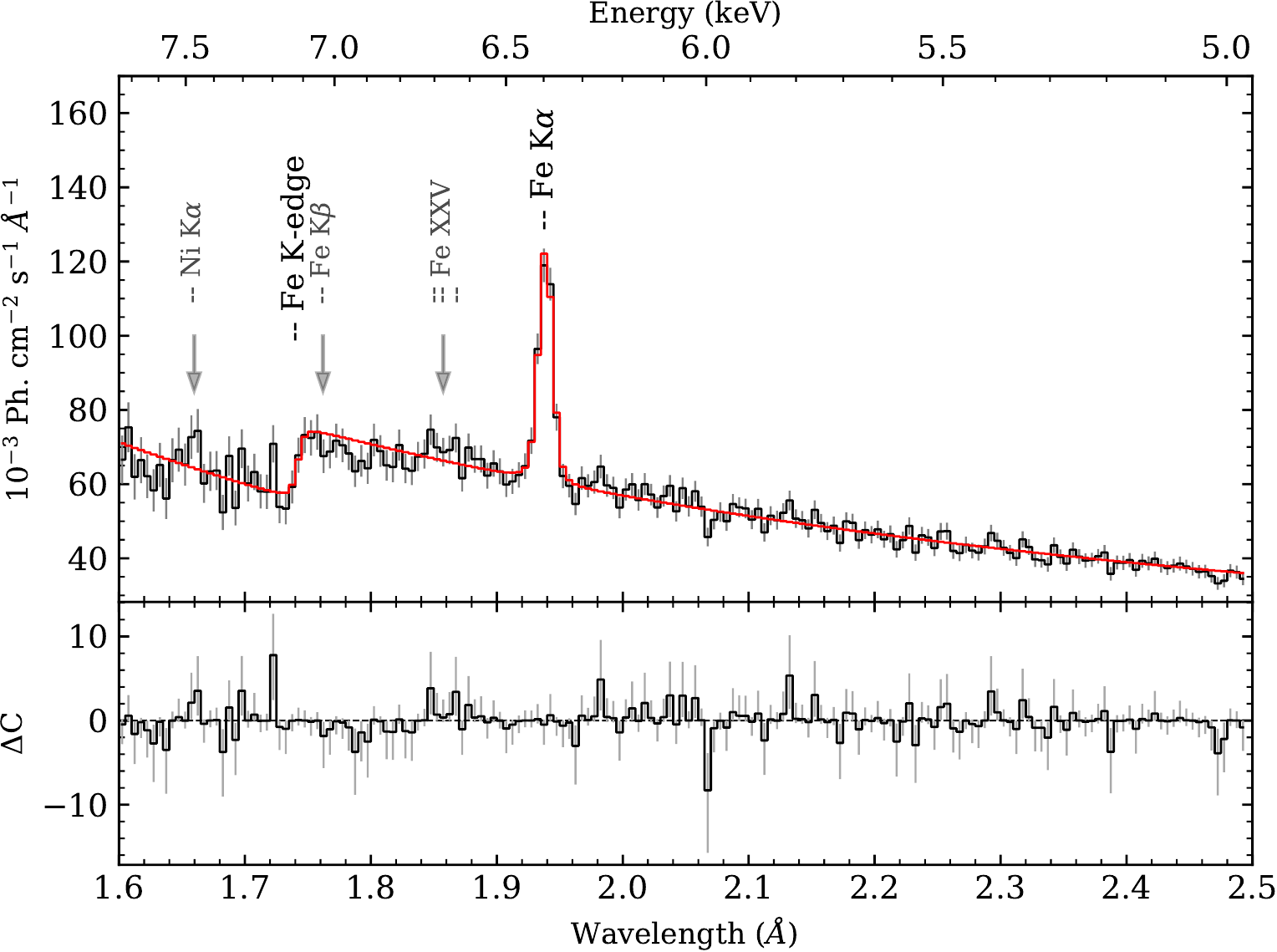}
      \caption{Fe-region spectrum and best-fit model (red line), with residuals shown in the bottom panel. The only line detected by the BB algorithm is identified and marked as a FeK$\alpha$ emission line, as well as the detected Fe K-edge. Arrows mark the position of the expected Ni K$\alpha$, Fe K$\beta$ and He-like \ion{Fe}{xxv} lines (in grey).}
     \label{fig:Fe}
   \end{figure}

\subsection{Sulphur region} 

We studied the S region in the wavelength range 4.5-6.0~\AA\ (Table \ref{tab:powerlaw_bestfit}). Line identification is based on the recent laboratory measurements from \citet{Hell2016}. The BB algorithm detected a single block between 5~\AA\ and 5.4~\AA, with $\alpha=27$. We model this block with the \ion{S}{xv} He-like triplet, the \ion{S}{xiv}, the \ion{S}{xi} and the blended fluorescence \ion{S}{ii-viii} lines. The second run of the algorithm resulted in the detection of the \ion{S}{xvi} Ly$\alpha$, with $\alpha=12$. Lastly, three more lines were detected: the \ion{Si}{xiii} He$\beta$ ($\alpha=8$), the \ion{S}{ix} ($\alpha=2.7$) and an unidentified absorption line at $\sim$5.457~\AA\ ($\alpha=2.2$). No reference wavelength was found for this last absorption line. Considering the low value of the parameter $\alpha$ and the lack of any other absorption feature in the whole spectrum, it is most likely that the line is just a statistical fluctuation. 

\begin{figure}
   \centering
   \includegraphics[width=\hsize]{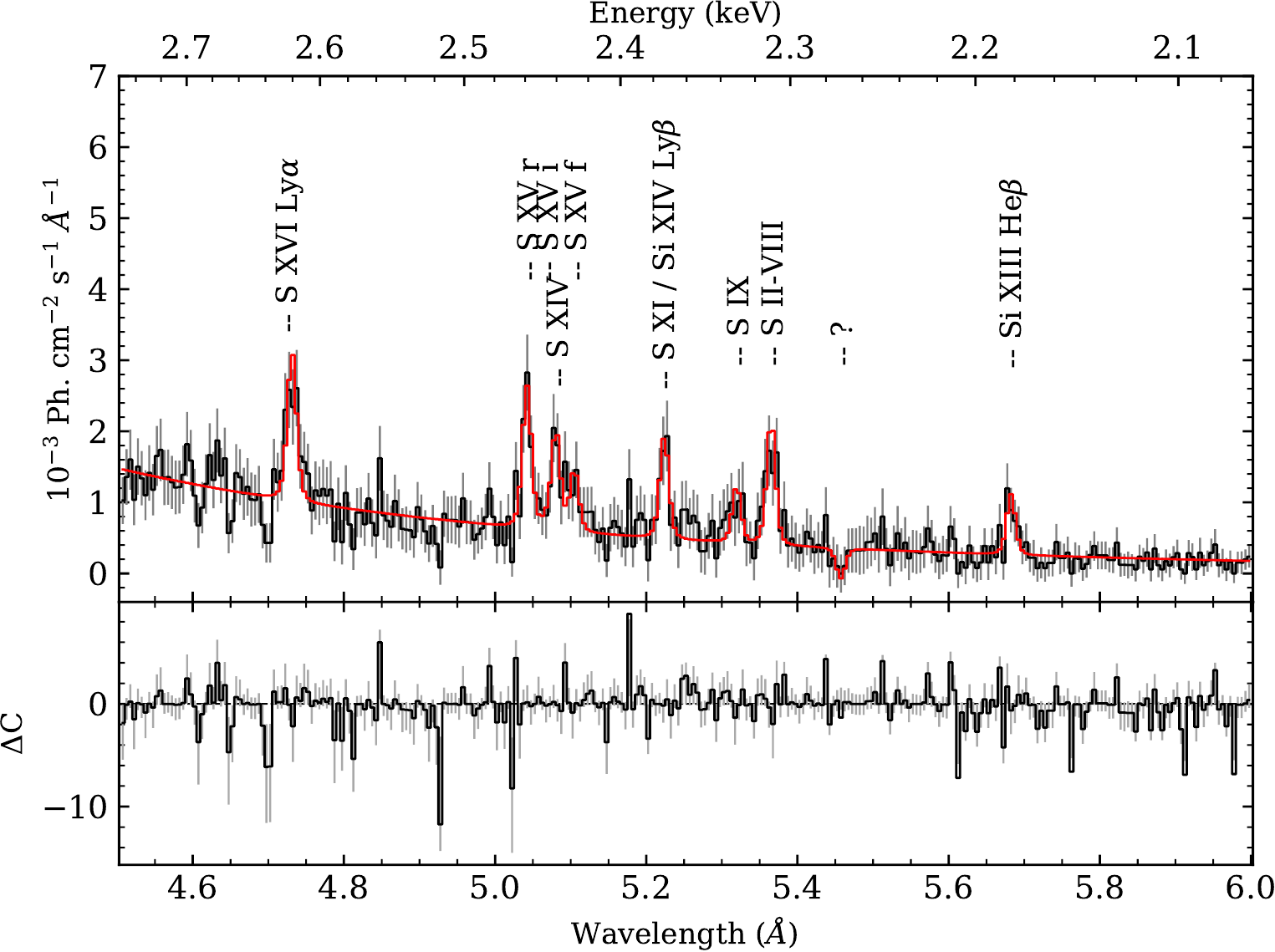}
      \caption{S-region spectrum and best-fit model (red line), with residuals shown in the bottom panel. The detected lines are labelled if identified.}
         \label{fig:S}
   \end{figure}

In the same region we could also expect to find the \ion{Si}{xiv} Ly$\beta$
line, at 5.217~\AA\ \citep{Erickson1977}. The lack of a significant detection of this line is probably due to the strong continuum. However, since the \ion{Si}{xiv} Ly$\alpha$ line is strong in the Si region (see Sect.\,\ref{sub:Si}), the \ion{Si}{xiv} Ly$\beta$ is most likely present and blended with the \ion{S}{xi} line. In Fig.\,\ref{fig:S}, we marked the line at 5.224 \AA\ with both its possible identifications. Given this line confusion, the Ly$\beta$/Ly$\alpha$ ratio for \ion{Si}{xiv} cannot be easily constrained. Only an upper limit of 0.55 can be derived, assuming the minimum flux for Ly$\alpha$ (cf. Sec.~\ref{sub:Si}) and that all flux of the discussed blend is due to \ion{Si}{xiv}~Ly$\beta$. Moreover, the high absorption constitutes a source of additional uncertainty as it influences the line ratio \citep{Kaastra_1995a}.

For this region, all the line widths were fixed to 0.003~\AA. Best-fit values are reported in Table \ref{tab:S}, together with the Doppler velocities computed with respect to laboratory reference values \citep{Hell2016}. Fig.\ref{fig:S} shows the spectrum, the best-fit model and the residuals of the fit. From the \ion{S}{xv} triplet, we obtained the best fit ratios $R=9.9^{+2.4}_{-2.2}$ and $G=0.48^{+0.14}_{-0.10}$ (Table \ref{tab:R_G_ratios}).

\begin{table*}
\renewcommand{\arraystretch}{1.3}
\caption{Spectral features detected in the S region. For each feature we report the detection parameter $\alpha$, the best-fit values (wavelength and line flux) and the Doppler velocities, computed using reference wavelengths measured by \citet{Hell2016}. Line widths fixed to 0.003~\AA\ for all the lines.}
\centering
\begin{tabular}{cccccc}
\hline\hline
Line & BB  & Ref. wavelength &  Det. wavelength & Line flux & Velocity\\
& $\alpha$ & (\AA) & (\AA) & (ph s$^{-1}$ cm$^{-2}$ $\times10^{-5}$) & (km s$^{-1}$)\\
\hline
\ion{S}{xvi} Ly$\alpha$ & 12 &4.7329\tablefootmark{a}& $4.731\pm0.003$ & $3.5^{+1.0}_{-0.9}$ & $-50^{+180}_{-170}$\\
\ion{S}{xv} $r$ & 27 & 5.0386 & $5.0422^{+0.0018}_{-0.0014}$ & $3.18^{+1.07}_{-0.99}$ & $210^{+110}_{-80}$ \\
\ion{S}{xv} $i$ & 27& 5.0666 & $5.0682^{+0.0018}_{-0.0014}$\tablefootmark{b} & $0.14\pm0.03$ & $=v_{(\text{\ion{S}{xv} r})}$ \\
\ion{S}{xv} $f$ & 27& 5.1013 & $5.1049^{+0.0018}_{-0.0014}$\tablefootmark{b} & $0.06^{+0.04}_{-0.07}$ & $=v_{(\text{\ion{S}{xv} r})}$\\
\ion{S}{xiv} &27& 5.0858 & $5.081\pm0.003$ & $2.3\pm0.8$ & $-310^{+180}_{-160}$ \\
\ion{S}{xi}/\ion{Si}{xiv} Ly$\beta$\tablefootmark{c} &27& 5.2250 & $5.224\pm0.002$ & $2.3^{+0.8}_{-0.7}$ & $-70\pm140$ \\
\ion{S}{ix} & 2.7 &5.3163& $5.320^{+0.006}_{-0.009}$ & $1.3^{+0.8}_{-0.7}$ & $210^{+340}_{-510}$\\
\ion{S}{ii-viii} &27& 5.3616 & $5.365\pm0.003$ & $2.8^{+0.9}_{-0.8}$ & $200\pm150$\\
& 2.2 &--& $5.457^{+0.002}_{-0.003}$ & $-0.69^{+0.011}_{-0.24}$ & --\\
\ion{Si}{xiii} He$\beta$ & 8 & 5.681\tablefootmark{d}& $5.683\pm0.003$ & $1.4^{+0.6}_{-0.5}$ & $80^{+150}_{-160}$\\
\hline\hline
\end{tabular}
\tablefoot{\citet{Hell2016} reports the statistical uncertainties for each energy, which correspond to an error in wavelength of the order of $10^{-4}--10^{-5}$~\AA. However, authors state that there is also a systematic uncertainty of 0.23 eV for S lines, which results in an error on the wavelength of 0.0008~\AA.
\tablefoottext{a}{\citet{GarciaMack1965}.}
\tablefoottext{b}{Distances between the $r$ line and the $i$ and $f$ lines computed from \citet{Drake1988}.}
\tablefoottext{c}{The reference wavelength of \ion{Si}{xiv} Ly$\beta$ is 5.217~\AA\ \citep{Erickson1977}. }
\tablefoottext{d}{\citet{Kelly1987}.}
}
\label{tab:S}
\end{table*}

\subsection{Silicon region}
\label{sub:Si}

We searched for Si lines in the region 6.0-7.4~\AA\ (Table \ref{tab:powerlaw_bestfit}). The BB algorithm highlighted at the first trial ($\alpha=190$) the \ion{Si}{xiv} Ly$\alpha$ line and a whole block in the range 6.6-6.8~\AA\ that we modelled with the He-like triplet \ion{Si}{xiii}, at first. The fluorescent line blend \ion{Si}{ii-vi} is detected with $\alpha=121$, while a whole block is detected at the wavelengths 6.9-7.1~\AA, with $\alpha=32$. We added three Gaussians to model this block, according to the laboratory measurements by \citet{Hell2016} \citep[see also][]{Grinberg_2017a}, corresponding to the \ion{Si}{vii}, \ion{Si}{viii} and \ion{Si}{ix} lines. The last detections are identified as the \ion{Al}{xiii} Ly$\alpha$ line ($\alpha=9$), the \ion{Si}{x} and \ion{Si}{xi} lines ($\alpha=5$) and the \ion{Si}{xii} line ($\alpha=1.8$). 

In the same region, also the RRC of \ion{Mg}{xii} is detected, at 6.321~\AA\ ($1.961\pm0.002$ keV), with a temperature of $4.5^{+5.8}_{-2.5}$~eV. Lastly, we added one more \mintinline{python}{redge} function to model the \ion{Mg}{xi} RRC, expected at 7.037~\AA\ \citep{Drake1988}. It results in a temperature of $3.1^{+1.6}_{-1.1}$ eV, consistent with the one of \ion{Mg}{xii} RRC (Table \ref{tab:RRC}).

\begin{figure}
   \centering
   \includegraphics[width=\hsize]{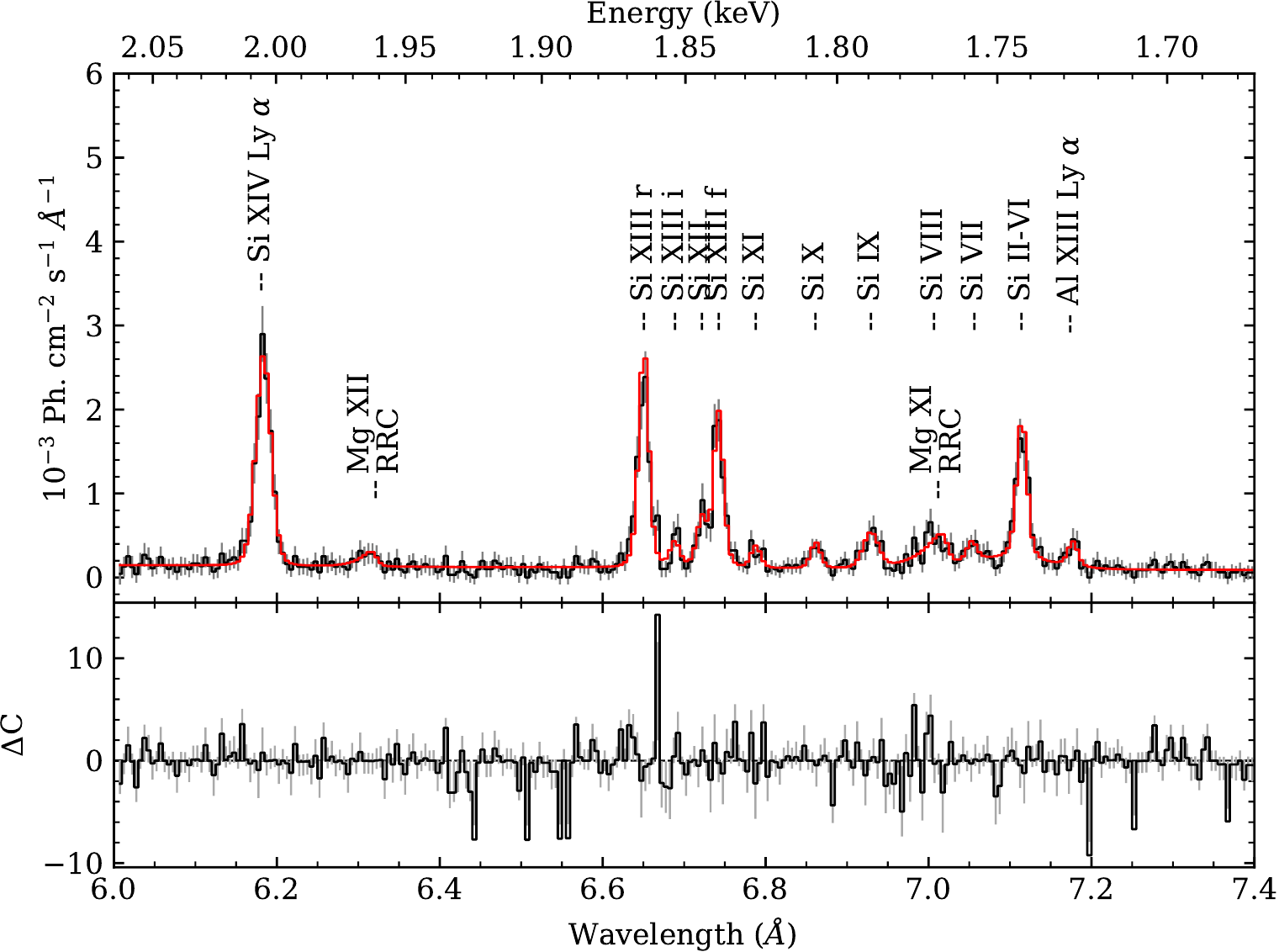}
      \caption{Si-region spectrum and best-fit model (red line), with residuals shown in the bottom panel. }
         \label{fig:Si}
   \end{figure}

The width of the lines was fixed to 0.003~\AA, except for the \ion{Si}{xiv} Ly$\alpha$ line, which has a slightly larger width of $\left(7.3^{+1.2}_{-1.1}\right)\times10^{-3}$~\AA. For each line, we computed the Doppler velocities with respect to the laboratory or literature reference wavelengths. All the best-fit values of the emission lines and RRCs are reported in Table \ref{tab:Si} and Table \ref{tab:RRC}, respectively, while the spectrum, the best-fit model and the residuals are shown in Fig.\,\ref{fig:Si}. The  best fit values of the $R$ and $G$ ratios of the \ion{S}{xiii} triplet resulted in $R=6.0\pm0.6$ and $G=0.80^{+0.10}_{-0.09}$ (Table \ref{tab:R_G_ratios}).
The BB algorithm did not detect the \ion{Mg}{xii} Ly$\beta$ emission line expected at $\sim7.1037$~\AA\  \citep{Erickson1977}. Also in this case, the line is most likely embedded in the (near-)neutral fluorescence \ion{Si}{ii-vi} lines.

\begin{table*}
\renewcommand{\arraystretch}{1.3}
\caption{Spectral features detected in the Si region. For each of them, we report the detection parameter $\alpha$, the best-fit values (wavelength and line flux). Line widths fixed to 0.003~\AA, if not stated otherwise. Doppler velocities of the Si lines, computed with respect to the reference wavelengths measured by \citet{Hell2016}. } 
\centering
\begin{tabular}{cccccc}
\hline\hline
Line & BB  & Ref. wavelength &  Det. wavelength & Line flux & Velocity\\
& $\alpha$ & (\AA) & (\AA) & (ph s$^{-1}$ cm$^{-2}$ $\times10^{-5}$) & (km s$^{-1}$)\\
\hline
\ion{Si}{xiv} Ly$\alpha$ & 190 & 6.1817\tablefootmark{a}& $6.184\pm0.001$ & $6.2\pm0.6$\tablefootmark{b} & $100\pm50$ \\
\ion{Si}{xiii} $r$ &190 & 6.6483 & $6.6506\pm0.0007$ & $4.5^{+0.8}_{-0.7}$ & $100\pm30$\\
\ion{Si}{xiii} $i$ &190 &6.7195& $6.6887\pm0.0007$\tablefootmark{c} & $0.51^{+0.05}_{-0.04}$ & $=v_{(\text{\ion{Si}{xiii} r})}$\\
\ion{Si}{xiii} $f$ &190& 6.7405 & $6.7427\pm0.0007$\tablefootmark{c} & $3.1\pm0.4$& $=v_{(\text{\ion{Si}{xiii} r})}$\\
\ion{Si}{xii} &1.8& 6.7197& $6.722\pm0.003$ & $0.9\pm0.3$ & $110\pm110$\\
\ion{Si}{xi} &5 &6.7841& $6.788\pm0.004$ & $0.42^{+0.18}_{-0.16}$ & $170\pm180$\\
\ion{Si}{x} &5& 6.8558& $6.862\pm0.004$ & $0.49^{+0.19}_{-0.17}$ & $270\pm180$\\
\ion{Si}{ix} &32& 6.9279 & $6.930\pm0.003$ & $1.1\pm0.3$ & $80\pm120$\\
\ion{Si}{viii} &32 &7.0008& $7.006\pm0.005$ & $1.6\pm0.3$ & 220$\pm$210\\
\ion{Si}{vii} &32& 7.0577& $7.057^{+0.005}_{-0.004}$ & $0.5\pm0.2$ & $-40^{+210}_{-170}$\\
\ion{Si}{ii-vi}\tablefootmark{d} &121& 7.1172& $7.115\pm0.001$ & $2.6\pm0.4$ & $-120\pm40$\\
\ion{Al}{xiii} Ly$\alpha$ &9 & 7.1764\tablefootmark{e} & $7.177\pm0.003$ & $0.6\pm0.2$ & $20^{+110}_{-120}$\\
\hline\hline
\end{tabular}
\tablefoot{
\citet{Hell2016} report  a systematic uncertainty of 0.13 eV for Si lines, corresponding to an error on the wavelength of 0.0005~\AA.\\
\tablefoottext{a}{\citet{GarciaMack1965}.}
\tablefoottext{b}{This line results in a best line width of $7.3^{+1.2}_{-1.1}\times10^{-3}$~\AA.}
\tablefoottext{c}{Distances between the $r$ line and the $i$ and $f$ lines computed from \citet{Drake1988}.}
\tablefoottext{d}{The Mg Ly$\beta$ \citep[7.1037~\AA,][]{Erickson1977} might be blended with the \ion{Si}{ii-vi} line.}
\tablefoottext{e}{\citet{Erickson1977}.}
}
\label{tab:Si}
\end{table*}

\subsection{Magnesium region}
\begin{figure}
   \centering
   \includegraphics[width=\hsize]{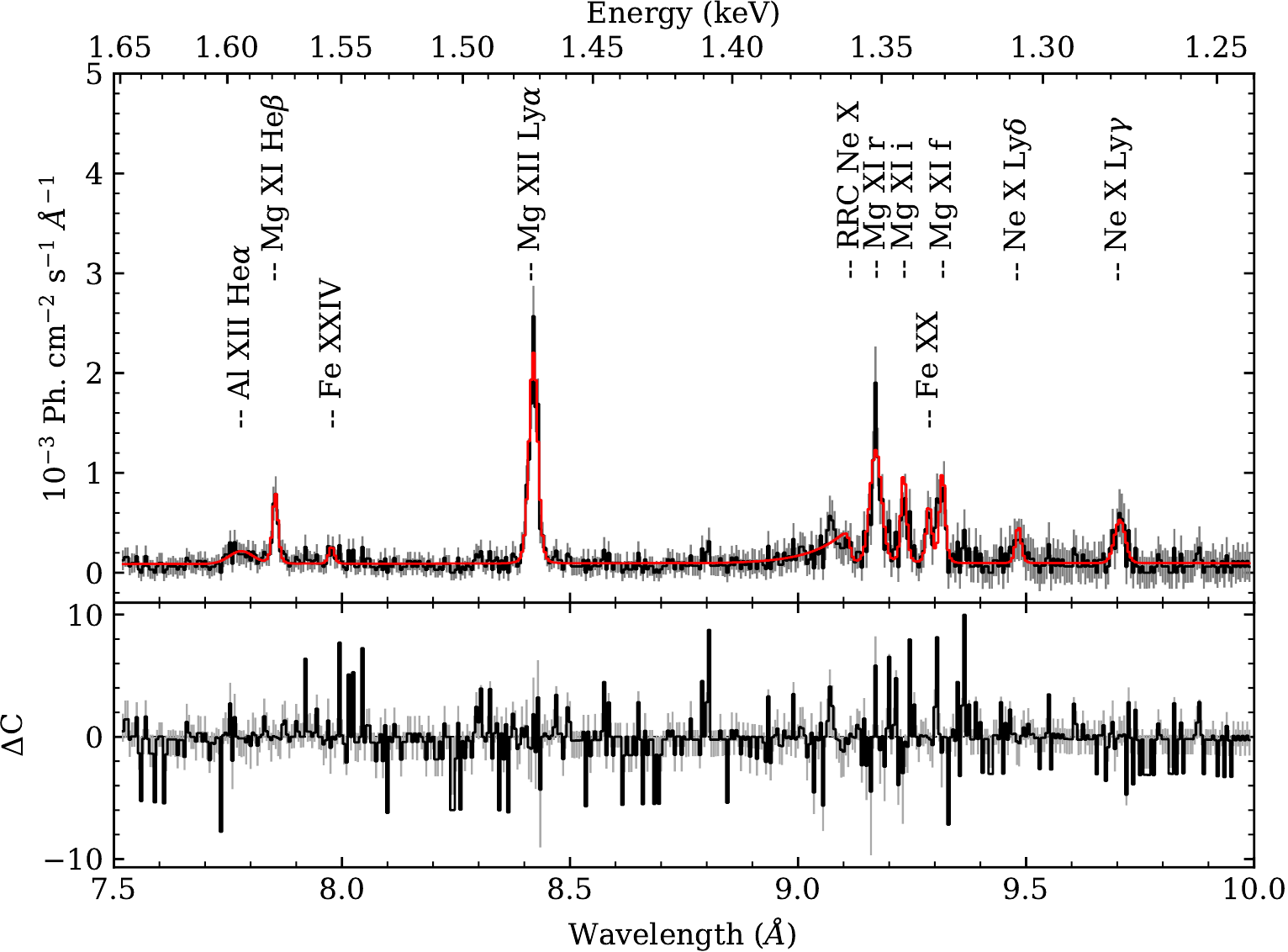}
      \caption{Mg-region spectrum and best-fit model (red line), with residuals shown in the bottom panel.}
         \label{fig:Mg}
   \end{figure}

The region we took into account to look for Mg emission lines ranges from 7.5~\AA\ to 10~\AA\ (Table \ref{tab:powerlaw_bestfit}). The first line detected corresponds to the \ion{Mg}{xii} Ly$\alpha$ ($\alpha=220$). The successive detection ($\alpha=89$) consisted in a block in the range $\sim$9-9.4~\AA, which we modelled with three Gaussians for the He-like triplet \ion{Mg}{xi}. In the same block, we insert the \ion{Ne}{x} RRC \citep{Schulz2002,Watanabe2006,Goldstein2004}. We also detected and identified the \ion{Mg}{xi} He$\beta$ ($\alpha=48$), the \ion{Ne}{x} Ly$\gamma$ ($\alpha=15$), the \ion{Al}{xii} $r$ He$\alpha$ ($\alpha=7$), the \ion{Ne}{x} He$\delta$ ($\alpha=4.4$), the \ion{Fe}{xx} ($\alpha=3.4$), and the \ion{Fe}{xxiv} ($\alpha=2.9$) emission lines. Best-fit value are reported in Table \ref{tab:Mg}, while the spectrum, the best-fit model and the residuals are shown in Fig.\,\ref{fig:Mg}. A few lines show a broadening that required to let their widths free. This is the case for \ion{Mg}{xii} Ly$\alpha$ whose width of $(7.4\pm1.2)\times10^{-3}$~\AA\ is in agreement with those of \ion{Si}{xiv} (Sect.\,\ref{sub:Si}) and \ion{Ne}{x} Ly$\alpha$ (Sect.\,\ref{sub:Ne}) lines. Other broadened lines are the \ion{Mg}{xi} $r$ and the \ion{Ne}{x} He$\delta$, $\sim$0.01~\AA\ width, and a \ion{Fe}{xxiii} line ($\sim$0.025~\AA\ width). The
\ion{Ne}{x} RRC, at a wavelength of $\sim$9.116~\AA\ ($1.3600^{+0.0012}_{-0.0010}$ keV) indicates a temperature of $10.8^{+3.4}_{-2.5}$ eV (Table \ref{tab:RRC}) consistent with previous findings at different orbital phases \citep{Schulz2002,Goldstein2004}. Doppler shifts of the Ly$\alpha$, the He$\beta$ and the triplet lines are around 150 \vel{}. From the intensities of the \ion{Mg}{xi} triplet we obtained the ratios $R=1.20^{+0.25}_{-0.23}$ and $G=0.74^{+0.13}_{-0.14}$ (Table \ref{tab:R_G_ratios}) for plasma diagnostic.

\begin{table*}
\renewcommand{\arraystretch}{1.3}
\caption{Spectral features detected in the Mg region (7.5--10~\AA). For each of them, we report the detection parameter $\alpha$, the best-fit values (wavelength and line flux) and the Doppler velocities, computed with respect to reference wavelength from literature. Line widths fixed to 0.003~\AA, if not stated otherwise.}
\centering
\begin{tabular}{cccccc}
\hline\hline
Line & BB  & Ref. wavelength &  Det. wavelength & Line flux & Velocity\\
& $\alpha$ & (\AA) & (\AA) & (ph s$^{-1}$ cm$^{-2}$ $\times10^{-5}$) & (km s$^{-1}$)\\
\hline
\ion{Al}{xii} He$\alpha$ & 7 & 7.7573\tablefootmark{a} & $7.782^{+0.012}_{-0.011}$ & $0.9\pm0.3$\tablefootmark{b} & $960^{+460}_{-430}$\\
\ion{Mg}{xi} He$\beta$& 48 & 7.850\tablefootmark{c} & $7.8565\pm0.0017$ & $1.2\pm0.3$ & $250\pm70$\\
\ion{Fe}{xxiv} & 2.9 & 7.985 \tablefootmark{d} & $7.980^{+0.008}_{-0.005}$ & $0.30^{+0.17}_{-0.14}$ & $-190^{+300}_{-190}$\\
\ion{Mg}{xii} Ly$\alpha$ & 220 & 8.42101\tablefootmark{e}& $8.4226\pm0.0011$ & $5.3^{+0.6}_{-0.5}$\tablefootmark{f} & $180\pm40$\\
\ion{Mg}{xi} $r$ & 89 & 9.16896\tablefootmark{a} & $9.1728\pm0.0015$ & $3.7\pm0.8$\tablefootmark{g} & $130\pm50$\\
\ion{Mg}{xi} $i$ & 89 & 9.2312\tablefootmark{a} & $9.2343\pm0.0015$\tablefootmark{a} & $1.51^{+0.14}_{-0.23}$ &$=v_{\text{(\ion{Mg}{xi} r)}}$ \\ 
\ion{Mg}{xi} $f$ & 89 & 9.3143\tablefootmark{a}& $9.3188\pm0.0015$\tablefootmark{a} & $1.5\pm0.4$ & $=v_{\text{(\ion{Mg}{xi} r)}}$ \\ 
\ion{Fe}{xx} \tablefootmark{h}& 3.4 & 9.282\tablefootmark{i}& $9.290\pm0.004$ & $1.0\pm0.4$ & $260\pm130$ \\
\ion{Ne}{x} Ly$\delta$ & 4.4 & 9.481\tablefootmark{e}& $9.485\pm0.006$ & $0.6\pm0.3$ & $130\pm190$\\
\ion{Ne}{x} Ly$\gamma$ & 15 & 9.708\tablefootmark{e}& $9.708\pm0.005$ & $1.3^{+0.5}_{-0.4}$\tablefootmark{j} & $0\pm150$ \\
\hline\hline
\end{tabular}
\tablefoot{
\tablefoottext{a}{\citet{Drake1988}.}
\tablefoottext{b}{Line width of $0.025^{+0.012}_{-0.008}$~\AA.}
\tablefoottext{c}{\citet{Kelly1987}.}
\tablefoottext{d}{\citet{Wargelin1998}.}
\tablefoottext{e}{\citet{Erickson1977}.}
\tablefoottext{f}{For this line the best-fit line width value was $\left(7.4\pm1.2\right)\times10^{-3}$~\AA.}
\tablefoottext{g}{Line width of $0.011\pm0.003$~\AA.}
\tablefoottext{h}{Close to the same wavelength there is also the \ion{Ne}{x} Ly$\zeta$ emission line at 9.291~\AA, but with a lower intensity ratio. In this case the resulting Doppler velocity would be ($-32\pm129$) km s$^{-1}$.}
\tablefoottext{i}{Unpublished atomic data from Liedahl (1997).}
\tablefoottext{j}{Line width of $0.010^{+0.005}_{-0.004}$~\AA.}
}
\label{tab:Mg}
\end{table*}


\subsection{Neon region}
\label{sub:Ne}
\begin{figure}
   \centering
   \includegraphics[width=\hsize]{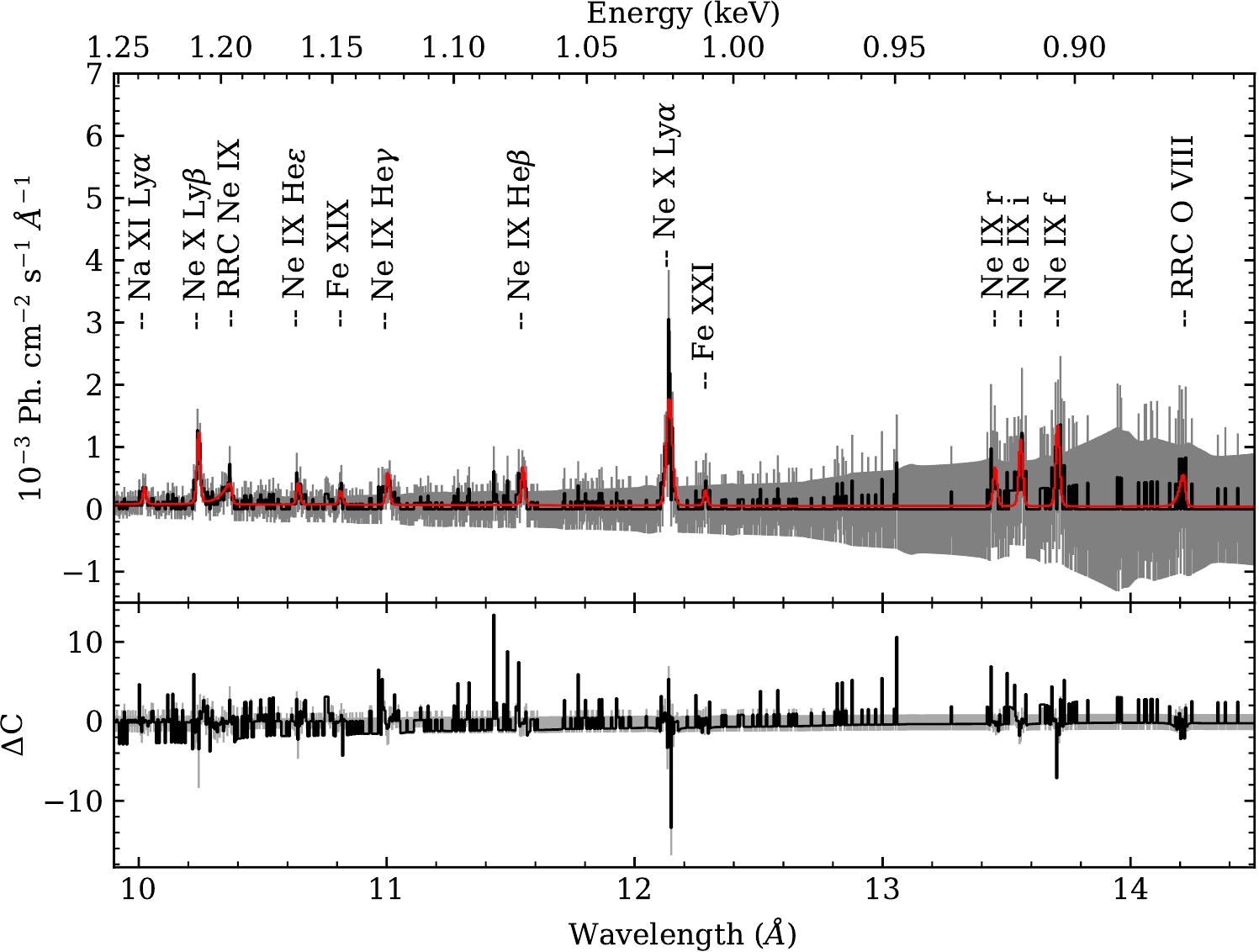}
      \caption{Ne-region spectrum and best-fit model (red line), with residuals shown in the bottom panel.}
         \label{fig:Ne}
   \end{figure}

The region for Ne emission lines goes from 10~\AA\ to 14.5~\AA\ (Table \ref{tab:powerlaw_bestfit}). We detected and identified 11 lines and two RRCs. Best-fit values are reported in Table \ref{tab:Ne} and \ref{tab:RRC}, the spectrum, best-fit model and residuals are shown in Fig.\,\ref{fig:Ne}. The first line to be detected by the BB procedure ($\alpha=49$)  was the \ion{Ne}{x} Ly$\alpha$, at a wavelength of 12.1398~\AA\ and with a width of $(9.6^{+3.0}_{-2.8})\times10^{-3}$~\AA. The successive detection ($\alpha=29$) was a line at $\sim$10.24~\AA, that we identified with the \ion{Ne}{x} Ly$\beta$. Hence, we fixed the distance of the latter line with respect to the corresponding Ly$\alpha$ according to \citet{Erickson1977}. The next detection ($\alpha=17$) was a block from 13.4~\AA\ to 13.9~\AA, that we modelled with the \ion{Ne}{ix} triplet \citep{Grinberg_2017a,Goldstein2004,Watanabe2006}. Lastly, we detected six more lines, corresponding to \ion{Ne}{ix} He$\beta$, at 11.549~\AA\ ($\alpha=8$), \ion{Ne}{ix} He$\gamma$ at 11.005~\AA\ ($\alpha=7$), \ion{Ne}{ix} He$\varepsilon$ at 10.644~\AA\ ($\alpha=3.2$), \ion{Na}{xi} Ly$\alpha$ at 10.023~\AA\ ($\alpha=2.5$), \ion{Fe}{xix} at 10.814~\AA\ ($\alpha=1.8$) and \ion{Fe}{xxi} at 12.285~\AA\ ($\alpha=1.7$). The \ion{Ne}{ix} RRC at 10.374~\AA\ 
was detected with $\alpha=8$ and resulted in a best-fit temperature of $4.5^{+3.4}_{-2.1}$ eV, while the \ion{O}{viii} RRC at 14.22~\AA\ was detected with $\alpha=2.8$ with a best-fit temperature of $0.9^{+4.2}_{-0.6}$ eV (Table \ref{tab:RRC}). This is the first detection of the \ion{O}{viii} RRC in \chandra\ data for Vela X-1. It was implied in ASCA observations \citep{Sako_1999a}, suggested by \citet{Schulz2002}, and only recently detected using \textsl{XMM-Newton} data \citep{Lomaeva2020}.  We note that the \ion{O}{viii} RRC might be also blended with a \ion{Fe}{xviii} line at 14.208~\AA\ \citep{Brown1998}.

We computed Doppler shifts for all the lines, obtaining velocities consistent with each other (Table \ref{tab:Ne}). The intensities of the lines of the \ion{Ne}{ix} triplet gave diagnostic best fit ratios of $R=1.2^{+0.6}_{-0.5}$ and $G=3.7^{+4.4}_{-1.7}$ (Table \ref{tab:R_G_ratios}).

\begin{table*}
\renewcommand{\arraystretch}{1.3}
\caption{Spectral features detected in the Mg region (10-14~\AA). For each of them, we report the detection parameter $\alpha$, the best-fit values (wavelength and line flux) and the Doppler velocities, computed with respect to reference wavelength from literature. Line widths fixed to 0.003~\AA, if not stated otherwise.}
\centering
\begin{tabular}{cccccc}
\hline\hline
Line & BB  & Ref. wavelength &  Det. wavelength & Line flux & Velocity\\
& $\alpha$ & (\AA) & (\AA) & (ph s$^{-1}$ cm$^{-2}$ $\times10^{-5}$) & (km s$^{-1}$)\\
\hline
\ion{Na}{xi} Ly$\alpha$\tablefootmark{a} &2.5& 10.023\tablefootmark{b} & $10.023\pm0.005$ & $0.5^{+0.3}_{-0.2}$ & $0\pm150$\\
\ion{Ne}{x} Ly$\beta$ &29& 10.23887\tablefootmark{c}& $10.2408\pm0.0017$\tablefootmark{d} & $2.1^{+0.6}_{-0.5}$& $60\pm50$\\
\ion{Ne}{ix} He$\varepsilon$\tablefootmark{e} &3.2&10.643\tablefootmark{b}& $10.644\pm0.006$ & $0.7^{+0.4}_{-0.3}$ & $30\pm170$\\
\ion{Fe}{xix} & 1.8 &$10.816$\tablefootmark{b}& $10.814^{+0.006}_{-0.005}$ & $0.4^{+0.3}_{-0.2}$ & $-60^{+170}_{-140}$\\
\ion{Ne}{ix} He$\gamma$\tablefootmark{f} &7& 11.001\tablefootmark{g} & $11.005^{+0.006}_{-0.007}$ & $1.0^{+0.5}_{-0.4}$ & $110^{+160}_{-190}$\\
\ion{Ne}{ix} He$\beta$\tablefootmark{h} &8 & 11.544\tablefootmark{g}& $11.549^{+0.005}_{-0.006}$ & $2.1^{+0.6}_{-0.5}$ & $130^{+130}_{-160}$\\
\ion{Ne}{x} Ly$\alpha$ & 49 & 12.132\tablefootmark{c}& $12.1398\pm0.0017$ & $5.3^{+1.2}_{-1.1}$ \tablefootmark{i} & $190\pm40$\\
\ion{Fe}{xxi} &1.7 & 12.284\tablefootmark{b} & $12.285^{+0.008}_{-0.007}$ & $0.5^{+0.5}_{-0.4}$ & $20^{+200}_{-170}$\\
\ion{Ne}{ix} $r$ &17& 13.4476\tablefootmark{j}& $13.454\pm0.005$&$1.3^{+1.6}_{-0.7}$& $140\pm110$\\
\ion{Ne}{ix} $i$ &17& 13.553\tablefootmark{j} & $13.557\pm0.005$ & $2.2^{+0.8}_{-0.6}$& $=v_{\text{(\ion{Ne}{ix} r)}}$\\ 
\ion{Ne}{ix} $f$ &17& 13.699\tablefootmark{j} & $13.706\pm0.005$ & $2.6^{+1.6}_{-1.3}$& $=v_{\text{(\ion{Ne}{ix} r)}}$\\ 
\hline\hline
\end{tabular}
\tablefoot{
\tablefoottext{a}{Possible line blends include \ion{Fe}{xx} at 10.024~\AA\ and \ion{Ni}{xxiv} at 10.027~\AA.}
\tablefoottext{b}{Reference wavelength taken from AtomDB database (\url{http://www.atomdb.org/index.php}).}
\tablefoottext{c}{\citet{Erickson1977}.}
\tablefoottext{d}{Computed from the \ion{Ne}{x} Ly$\alpha$ best fit wavelength, as from \citet{Erickson1977}.}
\tablefoottext{e}{Another possible identification is the \ion{Fe}{xix} at 10.648~\AA.}
\tablefoottext{f}{Possible line blends are: \ion{Fe}{xx} at 11.007~\AA, \ion{Na}{x} He$\iota$ at 11.003~\AA\ and \ion{Fe}{xix} at 11.002~\AA.}
\tablefoottext{g}{\citet{Kelly1987}.}
\tablefoottext{h}{Another possible line is \ion{Fe}{xx} at 11.546~\AA.}
\tablefoottext{i}{Best-fit value of the line width of $\left(9.6^{+3.0}_{-2.8}\right)\times10^{-3}$~\AA.}
\tablefoottext{j}{\citet{Drake1988}.}
}
\label{tab:Ne}
\end{table*}

\begin{table*}
\renewcommand{\arraystretch}{1.3}
\caption{Best-fit values of the RRCs in each region, with temperature reported in K and eV. The wavelengths in the last column are simply the conversion of the threshold energy from keV to~\AA\ and are meant just for convenience to the reader.}
\centering
\begin{tabular}{cccccc}
\hline\hline
RRC & Region & Threshold energy (keV) & Temperature ($10^4$ K) & Temperature (eV) & Wavelength (\AA)\\
\hline
\ion{Mg}{xii} & Si & $1.961\pm0.002$ & $5.2^{+6.7}_{-2.9}$ & $4.5^{+5.8}_{-2.5}$ & 6.321\\
\ion{Mg}{xi} & Si & $1.768\pm0.001$ & $3.6^{+1.9}_{-1.3}$ & $3.1^{+1.6}_{-1.1}$ & 7.022\\
\ion{Ne}{x} & Mg & $1.3600^{+0.0012}_{-0.0010}$ & $12.5^{+3.9}_{-2.9}$ & $10.8^{+3.4}_{-2.5}$ & 9.116\\
\ion{Ne}{ix} & Ne & $1.1950^{+0.0006}_{-0.0007}$ & $5.2^{+3.9}_{-2.4}$ & $4.5^{+3.4}_{-2.1}$ & 10.374\\
\ion{O}{viii} & Ne & $0.8720\pm0.0006$ & $1.0^{+4.8}_{-0.7}$ & $0.9^{+4.2}_{-0.6}$ & 14.218\\
\hline\hline
\end{tabular}
\label{tab:RRC}
\end{table*}

\begin{table*}
\renewcommand{\arraystretch}{1.3}
\caption{Best-fit values of the $G$ and $R$ ratios of the He-like triplets and correspondent electron temperatures and densities \citep{PorquetDubau2000}. The electron density of the He-like \ion{Si}{xiii} triplet (marked as $*$) is an upper limit.}
\centering
\begin{tabular}{cccccc}
\hline\hline
Element & $G$ & $R$ & Temperature (K) & Temperature (eV) & Electron density $n_\mathrm{e}$ (cm$^{-3}$)\\
\hline
\ion{S}{xv} & $0.48^{0.14}_{-0.10}$ & $9.9^{+2.4}_{-2.2}$ & -- & --\\
\ion{Si}{xiii} & $0.80^{+0.10}_{-0.09}$ & $6.0\pm0.6$ & $1\times10^7$ & 860 & $1\times10^{12}$\tablefootmark{*}\\
\ion{Mg}{xi} & $0.74^{+0.13}_{-0.14}$ & $1.2^{+0.3}_{-0.2}$ & $7\times10^6$ & 600 & $2\times10^{13}$  \\
\ion{Ne}{ix} & $3.7^{+4.4}_{-1.7}$ & $1.2^{+0.6}_{-0.5}$ & $1-3\times10^6$ & 90--260 & $1.5\times10^{12}$ \\
\hline\hline
\end{tabular}
\label{tab:R_G_ratios}
\end{table*}

\section{Photoionisation models with CLOUDY and SPEX}
\label{cloudy_spex_simulations}

We attempted a more physical modelling of the detected features using  photoionisation models with the latest release of CLOUDY \citep{Ferland2017,Chakraborty2020} and SPEX (v3.05, \citealt{kaastra1996_spex}, \citealt{kaastra2018_spex}). In both cases we used proto-Solar abundances from \citet{Lodders2009}.
Both these codes require an input ionising continuum. We approximated such a continuum with a sum of two components, as previously done in \citet{Grinberg_2017a} and \citet{Lomaeva2020}. The emission from the star, that dominates in the UV, was modelled with a black body, while the emission from the accretion onto the NS with a power law modified by a Fermi-Dirac cutoff. Both components have the same parameters as employed in \citet{Lomaeva2020}. 
In particular, the shape of the power law continuum cannot be well constrained at energies below 10\,keV, especially when strongly affected by absorption, as is the case with our observations. We thus used parameters derived from non-simultaneous \textsl{NuSTAR} observations \citep{Furst2014}. We note that there are some indirect hints that the illuminating continuum assumed here may not reflect the true continuum seen by the plasma in the system, such as, in particular, the large ratio between the Fe and Si/S fluorescence lines and the stability curves, which are unstable over wide ranges, especially at the ionisation parameters of interest. This emphasises the importance of strictly simultaneous observations at high resolution below 10\,keV and at energies above this range for the future.

In our modelling, we left free to vary the electron density $n_\mathrm{e}$ (cm$^{-3}$), the ionisation parameter $\xi$ (erg cm s$^{-1}$), the absorption coefficient $N_\mathrm{H}$ ($10^{22}$ cm$^{-2}$), and the turbulent velocity $v_\mathrm{turb}$ (\vel). We explored the parameter space with CLOUDY in the ranges $5.0\leq\log n_\mathrm{e}\leq11.5$, $0.0\leq\log\xi\leq4.0$, $20.9\leq\log N_\mathrm{H}\leq22.3$, and 80 \vel$\leq v_\mathrm{turb}\leq$160 \vel. For SPEX we assume a much larger parameter space since its PION model calculates the ionisation balance instantaneously and does not require a predefined grid of models.

We modelled the data with an absorbed partially covered power law, with spectral index $\Gamma=1$, corresponding to the input power law of our photoionisation models \citep{Furst2014}, in addition to the CLOUDY/SPEX photoionisation model. The absorption due to the interstellar medium was fixed to $3.7\times10^{21}$ cm$^{-2}$ \citep{HI4PI2016}, while the local absorption was left free to vary. The local partial absorption was applied only to the continuum, since both the geometry of the system, with localised wakes of material, and previous high resolution studies \citep{Schulz2002, Watanabe2006, Grinberg_2017a, Lomaeva2020} imply that the line producing region is not experiencing the same high absorption as the vicinity of the neutron star, where the continuum is produced. We further added three more Gaussians for the fluorescence Fe K$\alpha$ line, centred at 1.9388\, \AA\ (cfr. Sect.\,\ref{fig:Fe}), and for the near-neutral fluorescence emission lines of \ion{S}{ii-viii} and \ion{Si}{ii-vi}, which are not reproduced by CLOUDY and SPEX.  

\begin{figure*}
   \centering
   \includegraphics[width=.45\hsize]{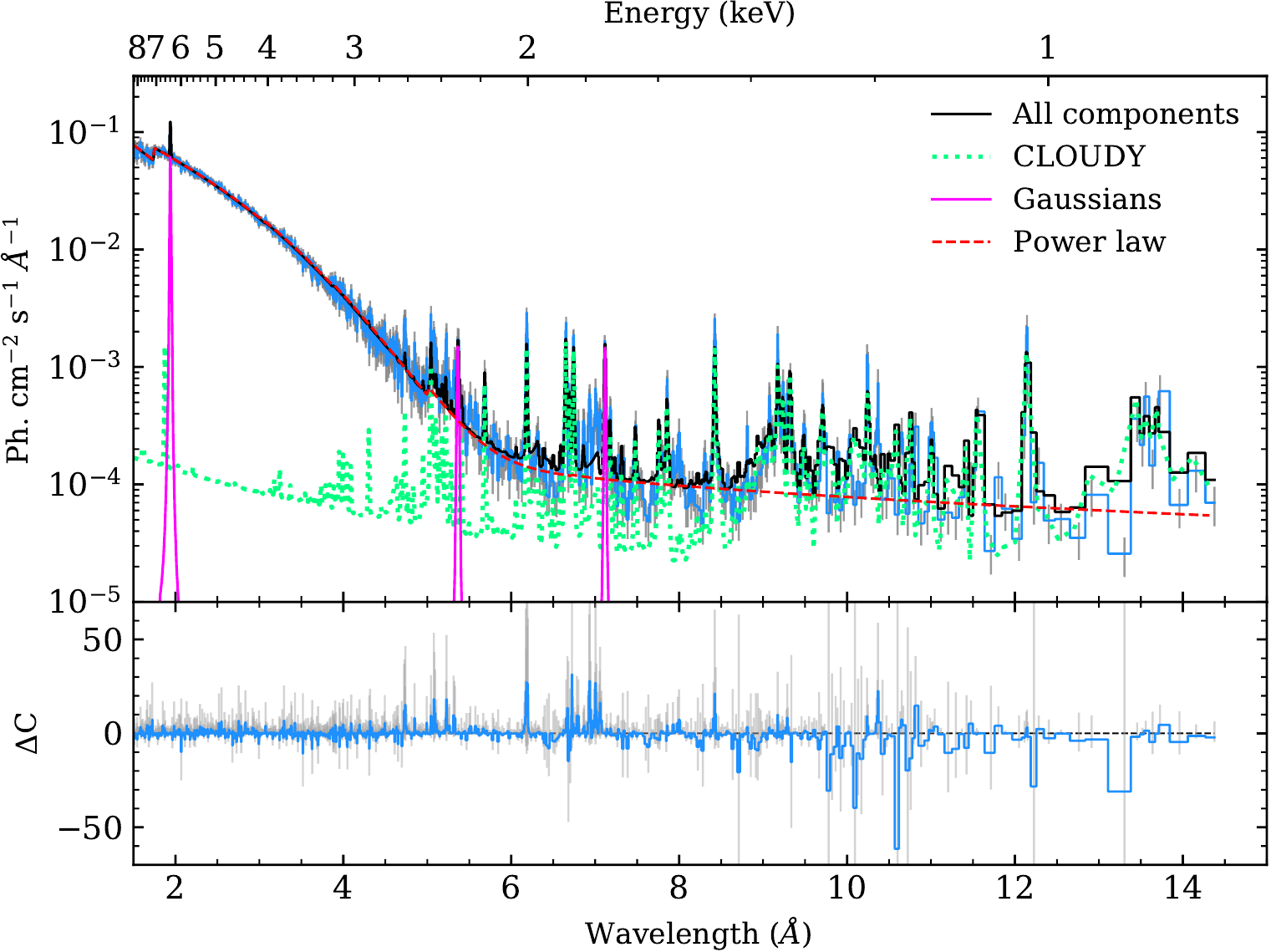} \quad
   \includegraphics[width=.45\hsize,height=.235\textheight]{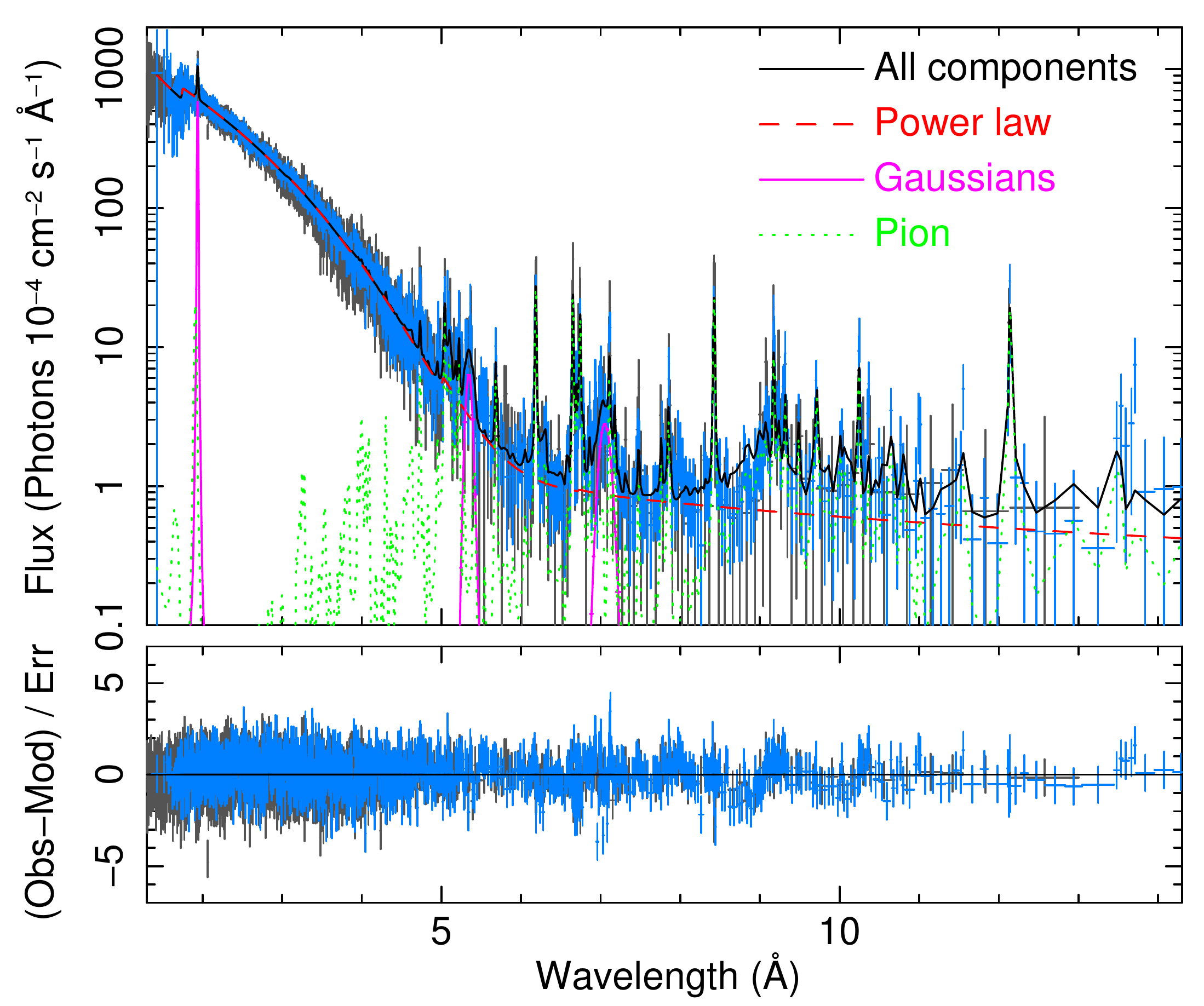}
      \caption{Fit of the whole spectrum with the photoionisation model (dotted green line) from CLOUDY (\textit{left panel}) and SPEX (\textit{right panel}), plus a partially covered power law (dotted red line), and three Gaussians for the fluorescence lines of Fe K$\alpha$, \ion{S}{ii-viii}, and \ion{Si}{ii-vi} (solid magenta lines). The total fit function is represented in black. Residuals of fit in the bottom panels. Spectra rebinned for clarity's sake. }
     \label{fig:cloudy_fit}
   \end{figure*}

The best fit CLOUDY model resulted in $\log n_\mathrm{e}=8.19984^{+0.00017}_{-0.01696}$, $\log\xi=3.728\pm0.009$, with $\log N_\mathrm{H}=22.175\pm0.020$~cm$^{-2}$ and a turbulent velocity of $\sim 160$ \vel. The model required a redshift, with a best fit value of $z\sim10^{-4}$, corresponding to a velocity of $v\sim100$ km s$^{-1}$, consistent with the Doppler shifts previously obtained. The Cash (d.o.f.) statistic value was 1.58 (2584). The modelling of the whole spectrum with SPEX resulted in the best fit values of $\log\xi=3.867^{+0.005}_{-0.009}$ and $N_H=(4.3\pm0.3)\times10^{21}$ cm$^{-2}$, with a line broadening of $160\pm16$~\vel\ and a Cash (d.o.f.) value of 1.57 (2382). Also in this case the model is redshifted with respect to the data, with a best fit velocity along the line of sight of $130^{+15}_{-20}$~\vel. Best fits are shown in Fig.\,\ref{fig:cloudy_fit}. We also tried to add a second CLOUDY component, obtaining no statistical significant improvement of the fit. 

We noticed that the electron density $n_\mathrm{e}$ is degenerate with the absorption of the interstellar medium (ISM): the larger the ISM $N_\mathrm{H}$, the larger the $n_\mathrm{e}$ (see discussion in Sect.\,\ref{sub:plasma_properties}).

\section{Discussion}
\label{sec:discussion}

We performed, for the first time, high-resolution spectroscopy
 analysis of \chandra/HETGS data of Vela X-1 at the orbital phase $\phi_{\text{orb}}\approx0.75$. A first look at the hardness ratio (Fig.\,\ref{fig:lcurve}) revealed no significant continuum spectral variability during the observation. The mainly flat shape of the hardness ratio is not surprising, since the line of sight at this orbital phase is expected to lie well within the photoionisation wake, a denser stream-like region that trails the NS \citep{Doroshenko2013,Malacaria2016} and acts as a constant absorber (see Fig.\ref{fig:sketch_velax1}).

The analysis pointed out the presence of Fe, S, Si, Mg and Ne, as well as of less intense emission lines from Al and Na. Contrary to previous observations \citep{Schulz2002,Goldstein2004,Watanabe2006}, there is no evidence of the presence of Ar ($\lambda\sim3.359$~\AA), Ca ($\lambda\sim4.186$~\AA) and Ni ($\lambda\sim1.660$~\AA) fluorescence lines. Upper limits of their fluxes resulted in $5.2\times10^{-5}$ ph s$^{-1}$ cm$^{-2}$ for Ar, $2.3\times10^{-5}$ ph s$^{-1}$ cm$^{-2}$ for Ca, and $3.1\times10^{-4}$ ph s$^{-1}$ cm$^{-2}$ for Ni, respectively.

In the next subsections, we discuss in details the Fe region (Sect.\,\ref{sub:Fe_region}), carry out plasma diagnostic (Sect.\,\ref{sub:plasma_properties}), and investigate the geometry of the wind of the companion star (Sect.\,\ref{sub:wind_geometry}).

\subsection{The Iron complex}
\label{sub:Fe_region}

The Fe region (1.6--2.5 \AA) is dominated by a Fe K$\alpha$ line, centred at $1.9388\pm0.0006$ \AA. Assuming no Doppler shift for the line, the corresponding maximum ionisation state is \ion{Fe}{x} \citep{Palmeri2003}, consistent with the results of \citet{Grinberg_2017a} (below \ion{Fe}{xii}), and different from the case of an irradiated wind, as showed by the hydrodynamic simulations of \citet{Sander2018} (where the wind is mainly driven by \ion{Fe}{iii} ions). However, the line may be redshifted so that a higher ionisation state could be expected. A more refined calculation is beyond the goal of this paper. 

The only other relevant feature detected in this region is the Fe K-edge at $1.742\pm0.003$ \AA\ (see Table \ref{tab:Fe}), which is not significantly Doppler shifted.

The BB algorithm did not detect the Fe K$\beta$ line, expected at $\sim$1.758~\AA, most likely because of the proximity of the Fe K-edge.  However, since the average flux ratio between the Fe K$\beta$ and Fe K$\alpha$ lines is 0.13--0.14 \citep[for the charge states \ion{Fe}{II-IX}]{Palmeri2003}, we can estimate an expected flux of $(1.32\pm0.11)\times10^{-4}$ \flux, which might not be sufficient to let the line emerge from the continuum underneath. To verify this assertion, we generated 1000 Monte Carlo simulated spectra adding to the best fit model a Gaussian at the correspondent wavelength of the Fe K$\beta$ with the expected flux. We then run the BB algorithm on all the simulated spectra (cf. Sect.\,\ref{sec:iron}). In no case the line was detected, confirming its weakness with respect to the X-ray continuum and the K-edge, which precluded a detection in the observational data. The Fe K$\beta$/K$\alpha$ ratio depends on the ionisation of iron \citep[see the detailed discussions in][]{Molendi2003,Bianchi2005}. For higher charge states, the expected line ratio is even smaller, i.e., the Fe K$\beta$ line would be even weaker than what our simulation showed as undetectable. Therefore, we cannot rule out that the ionisation state is higher than what we assumed.
 We discuss the prospects of detecting Fe K$\beta$ with future instruments in Sect.\,\ref{athena_xrism_simulations}.

Results from \citet{Goldstein2004} at ${\phi_\mathrm{orb}\approx0}$ and ${\phi_\mathrm{orb}\approx0.5}$ show, in the same spectral region, the presence of the Ni Ly$\alpha$ line at $\lambda\sim1.660$ \AA, while \citet{Schulz2002} propose the presence of a \ion{Fe}{xxv} emission line at $\lambda\sim1.85$ \AA\ ($\phi_\mathrm{orb}\approx0$). The BB procedure did not detect any feature at those wavelengths, but after a visual inspection, we noted a marginal presence of residuals in emission. So we add two more Gaussians to the best fit model of the Fe region, at $\lambda\sim1.66$ \AA\ and $\lambda\sim1.86$ \AA, for the Ni Ly$\alpha$ and a \ion{Fe}{xxv} respectively, and fit the spectrum again. The Fe XXV is actually a He-like triplet, but the resolution of the MEG of 0.023 \AA\ FWHM, adopted consistently through the paper, is not good enough to resolve the lines individually. 
Hence, we use just one Gaussian to fit the whole ion, letting the width free to vary. The width of the Ni Ly$\alpha$ line was fixed to the usual value of 0.003~\AA. The fluxes of these latter Gaussians resulted in $\left(1.8^{+1.3}_{-1.2}\right)\times10^{-4}$ \flux for the Ni Ly$\alpha$ and $\left(3.1\pm1.2\right)\times10^{-4}$ \flux for the \ion{Fe}{xxv} lines, while the width of the He-like \ion{Fe}{xxv} had a best fit value of $0.018^{+0.013}_{-0.007}$ \AA.

From the Fe edge (Table \ref{tab:Fe}), we computed the equivalent hydrogen column as ${N_\mathrm{H}=\tau_\mathrm{edge}/(Z_\mathrm{Fe}\sigma_\mathrm{Fe})}$, where ${Z_\mathrm{Fe}=2.69\times10^{-5}}$ is the solar Fe abundance \citep{wilms2000} and ${\sigma_\mathrm{Fe}=3.4\times10^{-20}}$ cm$^2$ is the photoelectric absorption cross section for \ion{Fe}{xxv} at the wavelength of the K-edge \citep{Verner1996}. Using the best-fit value optical depth ${\tau_\mathrm{edge}=0.31\pm0.03}$, we derive ${N_\mathrm{H}=(3.4\pm0.3)\times10^{23}}$ cm$^{-2}$, which is nearly consistent with the best-fit value of ${N_\mathrm{H}=(2.68\pm0.07)\times10^{23}}$ cm$^{-2}$ obtained fitting the spectrum in this region with a simple absorbed power law, with solar abundances and cross sections as specified in Sect.\,\ref{sec:high_resolution_spectroscopy}. These values are of the same order of magnitude as the best fit values found for observations using \textit{MAXI} \citep{maxi2009} by \citet{Doroshenko2013} and \textit{NuSTAR} \citep{nustar2013} by \citet{Furst2014} at the same orbital period. However, we must bear in mind here that the model we used does not account for the Fe K$\beta$ line, which may contribute to larger uncertainties on the Fe K-edge parameters.

\subsection{Plasma properties}
\label{sub:plasma_properties}

The presence of five narrow RRCs (\ion{Mg}{xi}, \ion{Mg}{xii}, \ion{Ne}{ix}, \ion{Ne}{x}, and \ion{O}{viii}) suggests that the plasma is photoionised, with a temperature between $\sim$1 and 10 eV (Table \ref{tab:RRC}).
A further indication of a photoionised plasma might be the value of $G=3.7^{+4.4}_{-1.7}$ of the \ion{Ne}{ix} triplet (Table \ref{tab:R_G_ratios}), consistent with 4 in spite of the large uncertainties \citep{PorquetDubau2000}. 

However, the $G$ ratios of \ion{S}{xv} ($G=0.48^{+0.14}_{-0.10}$), \ion{Si}{xiii} ($G=0.80^{+0.10}_{-0.09}$) and \ion{Mg}{xi} ($G=0.74^{+0.13}_{-0.14}$) are all smaller than 1, indicating that collisional processes are not negligible and may even dominate \citep{PorquetDubau2000,Porquet2010}.  
Under the hypothesis of a collisional equilibrium plasma (CIE), we can estimate the temperature from the $G$ ratio values \citep{PorquetDubau2000}.
From the He-like \ion{Si}{xiii} and \ion{Mg}{xi} triplets we obtain temperatures of $\sim1\times10^7$~K and $\sim7\times10^6$~K, respectively, which are two orders of magnitude higher than the ones from the Ne RRCs. 

This inconsistency between temperatures derived from the RRCs and the He-like line ratios is likely due the known issue that relative level populations between the upper levels of the He-like triplet lines can be shifted by other physical phenomena, which are likely present in HMXBs, thus making the $G$ ratio unreliable.
In particular, two processes can enhance a resonant $r$ line stronger than the intercombination $i$ or forbidden $f$ lines: photoexcitation and resonance line scatter. Photoexcitation can be important in photoionisation equilibrium (PIE) plasma, when many photons with the right energy excite the electrons to the resonant level. This clearly enhances the resonance line and, then, alters the $G$ ratio with respect to the pure recombination case \citep[see the comprehensive explanation in][]{Kinkhabwala2002}. The presence of a few weak iron L emission lines (\ion{Fe}{xix-xxiv}) also seems to point in this direction \citep{Sako2000}. 

Resonant line scattering occurs when a photon is absorbed and re-emitted in the same wavelength, but in the direction of the lowest optical depth. 
This phenomenon is well explained by \citet{Wojdowski2003} for the HMXB Centaurus X-3, observed during eclipse. In the case of Vela X-1, though we are not in the eclipsing phase, the dense streams of matter surrounding the NS can act as a strong absorber, enhancing the resonance line scattering into the line of sight. 

Concerning the $R$ ratio, the values of \ion{Mg}{xi} ($R=1.2^{+0.3}_{-0.2}$) and \ion{Ne}{ix} ($R=1.2^{+0.6}_{-0.5}$) He-like lines implies an electron density of the plasma of $\sim2\times10^{13}$ cm$^{-3}$ and $\sim1.5\times10^{12}$ cm$^{-3}$, respectively, considering a plasma temperature of $7\times10^6$ K and $2\times10^6$ K, as previously estimated\footnote{We note here that the $R$ ratio depends upon  the relative ionic abundance of the H-like and He-like ions ($\chi_\text{ion}$ parameter), but in the range of our interest the dependence is so small that we can neglect it \citep[see Fig.\,9 of][]{PorquetDubau2000}.}. On the other hand, the $R$ ratios of \ion{Si}{xiii} ($R=6.0\pm0.6$) and \ion{S}{xv} ($R=9.9^{+2.4}_{-2.2}$) are much higher than the respective values at the low density limit, when the relative intensities of the He-like lines are in fact independent of the electron density of the plasma. In the case of Si, for instance, the low density limit value is $R=3$, corresponding to a maximum density of the order of $10^{12}$\,cm$^{-3}$  \citep{PorquetDubau2000}, which can be addressed here as upper limit. On the other hand, the fit with CLOUDY and SPEX photoionisation models highlighted the degeneracy of the electron density $n_\mathrm{e}$ with the model chosen for the continuum, and, in particular, with the absorption from the ISM. The best fit value of  $n_\mathrm{e}=1.5\times10^{8}$\,cm$^{-3}$, for instance, can be treated only as a lower limit. 
The analysis underlines that the estimation of the density is influenced in opposite directions by the $R$ ratio and the continuum and the real value is somewhere in between those limits.  

Also the UV radiation of the companion star can alter the plasma \citep[the so-called ‘‘UV-pumping'' mechanism,][]{GabrielJordan1969, Blumenthal_1972a, Mewe_1978a,Porquet_2001a}. UV radiation mimics a high density plasma, favouring the population of the $^3P$ levels against the $^3S_1$ level, leading to an increase of the intensity of the intercombination line, against the forbidden line and, hence, to smaller values of the $R$ ratio. The influence of the UV emission is taken into account in both, CLOUDY and SPEX based photoionisation models, through our choice of the continuum. Such models should also, if applicable to the given data at all, give better constrains on the underlying plasma parameters than the more empirical consideration of $G$ and $R$ ratios. The quality of our fits in Sect.\,\ref{cloudy_spex_simulations} imply that this is the case.

Overall, both the self-consistent photoionisation codes provided a satisfactory fit of the data (Fig.\,\ref{fig:cloudy_fit}), implying that, at this specific orbital phase, the plasma is mainly photoionised. However, a closer inspection at the residuals hints to the presence of at least another phase of the plasma. The near-neutral emission lines of \ion{S}{ii-viii} and \ion{Si}{ii-vi}, as well as the Fe K$\alpha$ line are not reproduce by the photoionisation models that are driven by the presence of highly ionised lines. This naturally suggests that the plasma cannot be a single component plasma. 

In a possible scenario, colder and denser clumps of plasma, from either the wind or larger scale accretion structures such as wakes, can cross unevenly the line of sight, adding to the PIE emission of the wind of the companion star a further component with a lower ionisation. Our data do not allow to constrain the origin of this component that could be, for example, a further, colder PIE component, a collisionally ionised component or a more complex mix with a temperature gradient as is the case, e.g., in Cyg X-1 \citep{Hirsch2019}. We also note that our results emphasise the necessity of an accurate treatment of intermediate and low ionisation ions in atomic codes used for high resolution X-ray spectroscopy.

\begin{figure}
   \centering
   \includegraphics[width=\hsize]{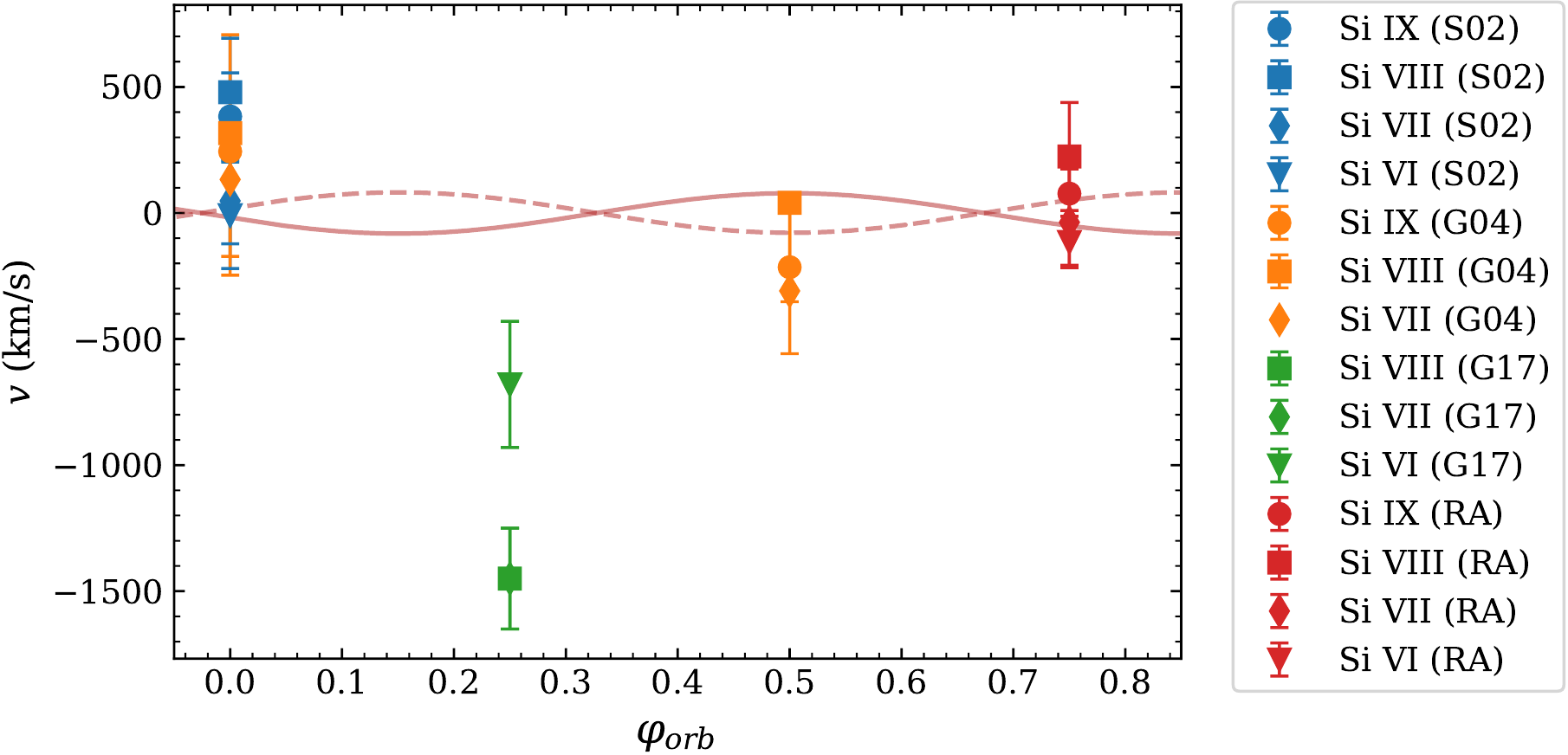}
      \caption{Doppler velocities at different orbital phases of \ion{Si}{ix} (circles),  \ion{Si}{viii} (squares), \ion{Si}{vii} (diamonds) and \ion{Si}{vi} (reverse triangles), from \citet{Schulz2002} (S02, \textit{blue}) and \citet{Goldstein2004} (G04, \textit{orange}), as adjusted for laboratory reference values by \citet{Hell2016}, from \citet{Grinberg_2017a} (G17, \textit{green}) and from the present work (RA, \textit{red}). The solid and dashed lines stand for the radial velocities of the NS and the giant star, respectively.}
         \label{fig:velocities_Si}
   \end{figure}

\begin{figure}
   \centering
   \includegraphics[width=\hsize]{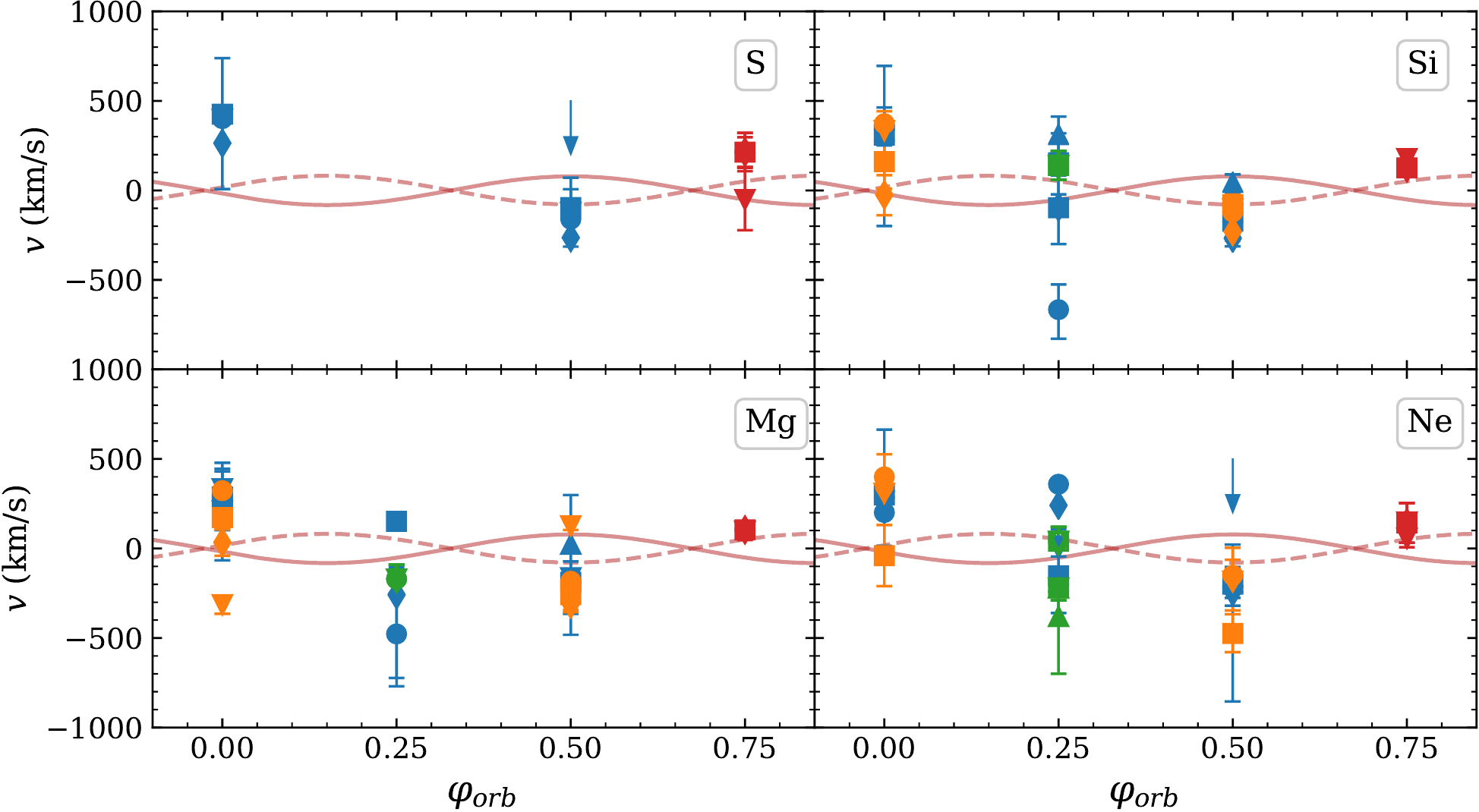}
      \caption{Doppler velocities at different orbital phases of Ly$\alpha$ lines and He-like triplets of S, Si, Mg and Ne from \citet{Schulz2002} (\textit{blue}) and \citet{Goldstein2004} (\textit{orange}), from \citet{Grinberg_2017a} (\textit{green}) and from the present work (\textit{red}). Different symbols stand for different ionisation stages. The solid and dashed lines represent the radial velocities of the NS and of the companion star, respectively.}
     \label{fig:velocities_all}
   \end{figure}
   
\subsection{Wind geometry}
\label{sub:wind_geometry}
Doppler velocities at different orbital phases can reveal the location and dynamics of the line emitting material. Fig.\,\ref{fig:velocities_Si} shows the velocities for the ions of \ion{Si}{vi-ix} from \citet{Schulz2002} and \citet{Goldstein2004} at the orbital phases $\phi_\mathrm{orb}\approx0$ and $\phi_\mathrm{orb}\approx0.5$, adjusted with respect to the laboratory measurements of \citet{Hell2016}, together with the ones from \citet{Grinberg_2017a} at the orbital phase $\phi_\mathrm{orb}\approx0.25$, and with those in the present work ($\phi_\mathrm{orb}\approx0.75$). Velocities at $\phi_\mathrm{orb}\approx0.25$ are negative (blueshift), while velocities at the other orbital phases are positive (redshift) and/or consistent with no shift. The same behaviour is observed also for all the others lines of S, Si, Mg and Ne (Fig.\,\ref{fig:velocities_all}), even though there are no recent laboratory measurements that allow us to validate the Doppler shifts found by the previous studies \citep{Schulz2002,Watanabe2006,Goldstein2004,Grinberg_2017a}. 
Most of the velocities are consistent with the radial velocity of the NS, as well as of the companion star (solid and dashed lines in Figs.\,\ref{fig:velocities_Si}-\ref{fig:velocities_all}), computed as:
\begin{equation}
v_{rad}=2\pi a\sin{i}[\cos{(\vartheta+\omega)}+e\cos{\omega}]/(T\sqrt{1-e^2})
\end{equation}
where $a$ is the semi-major axis, $i$ is the inclination, $T$ is the orbital period, $e$ is the eccentricity, $\vartheta$ and $\omega$ are the true anomaly and the argument of periapsis, respectively.

The overall behaviour is consistent with the material co-moving with the NS, though the lack of more observational data for each orbital phase prevent us to assert it definitively. However, this behaviour has already been observed for the black hole HMXB Cygnus X-1 \citep{Hirsch2019,Miskovicova2016}, where the Doppler shifts show a clear modulation with the orbital phase. It has already been suggested for Vela X-1 that the wind velocity at the distance of the NS is  $\sim$100 \vel{}
and lower than
typically estimated from prescribed simple $\beta$-laws \citep{Sander2018}. The large spread in the range of observed Doppler shifts within the same orbital phases may be due radiation coming from regions further downstream the wind or due to a more complex velocity structure in the accretion region. 

More considerations on the geometry of the emitting region can be drawn from the ionisation state of the plasma. The ionisation parameter can be expressed as in \citet{Tarter1969}:
\begin{equation}
    \xi=\frac{L_X}{n_\mathrm{e}r^2}
\end{equation}
where $L_X$ is the X-ray luminosity of the source, $n_\mathrm{e}$ the particle density of the plasma, and $r$ the distance at which the lines are produced. From the photoionisation models, we computed the distributions of the relative abundances of all the ions as a function of $\xi$. For the H-like ions, the ionisation parameter was in the range $3.7\geq\log(\xi)\geq4.2$. Assuming that each ion is produced at the peak of its distribution, for a luminosity of $4\times10^{36}$ \ergs and a best-fit value of $\log(n_\mathrm{e})\sim8$, we obtained a distance in the range $(1.6-2.8)\times10^{12}$ cm $= 23-40\ R_\odot$.
Considering that the orbital separation of the system is $\sim$50 $R_\odot$ and the companion star has a radius of about 30 $R_\odot$, the region where the H-like emission lines come from seems to be very close to the surface of the companion star, rather then to the surface of the NS. The other ionisation stages have lower values of $\log(\xi)$, implying even higher distances, compatible with the idea of a wake expanding after the passage of the NS. We note, however, that this estimate assumes a constant density, that is most likely not the case for an expanding wind, even without taking into account possible clumping and wake structures.

Our result is in agreement with simulations of X-ray photons in a smooth wind from \citet{Watanabe2006}, for H-like Si, in the case of a mass loss rate $\dot{M}\leq1.0\times10^{-6}\ M_\odot$yr$^{-1}$, consistent with the latest estimation for Vela X-1 of $\sim0.7\times10^{-6} M_\odot$ yr$^{-1}$ \citep{Sander2018}. In the end, our simple calculation would suggest that the photoionised plasma is produced at the orbital separation of the system, in a region close to the surface of the companion star.

Nonetheless, given the uncertainty in the electron density driven by the continuum, we repeated the calculation using instead the $n_\mathrm{e}$ derived from the $R$ ratio ($n_\mathrm{e}\sim10^{12}$ cm$^{-3}$). This $n_e$ value, however, does not take into account the presence of the strong UV radiation from the stellar wind (Sec.~\ref{sub:plasma_properties}) and thus has to be considered as an overestimate. For the same values of the ionisation parameter as before, the resulting distance is $r\lesssim0.5\,R_\sun$, comparable with the Bondi-Hoyle-Littleton radius of the NS in Vela X-1 of $\sim10^{10}$ cm \citep{Manousakis2015}. The assumption of using the same ionisation parameters holds because $\log(\xi)$ is primarily driven by the ionisation state and thus hardly changes with the electron density, which is instead driven by the absolute line strength (i.e., distance and continuum) and the triplet shape, if the lines are well resolved. In this case, of course, the ionisation of the wind would be due almost entirely to the gravitational pulling of the NS. 

From this analysis, we cannot infer the presence of clumps.

\section{Future perspectives with \textit{XRISM}/Resolve and \textit{Athena}/X-IFU}
\label{athena_xrism_simulations}

High-resolution spectroscopy is a powerful tool to study X-ray emission from any kind of astrophysical plasma. Currently, limitations of X-ray satellites are due, for instance, to their intrinsic resolution and sensibility. 
New generation X-ray satellites will go beyond these limits. The X-Ray Imaging and Spectroscopy Mission \citep[\xrism, formerly \textit{XARM},][]{Tashiro2018} and the Advanced Telescope for High Energy Astrophysics \citep[\athena,][]{Nandra2013} will host on-board microcalorimeters with an energy resolution down to a few eV, thus exceeding the
resolution of Chandra gratings in the Fe K region.

We performed simulations of this region (1.6--2.2 \AA, cfr. Sect.\,\ref{sub:Fe_region}), including the Fe K-edge and the Fe K$\alpha$ as detected in the \chandra\ observation, and the Fe K$\beta$, the He-like \ion{Fe}{xxv} and the Ni K$\alpha$ with the upper limit on the flux as in Sect.\,\ref{sub:Fe_region}. Both microcalorimeters should be able to resolve the Fe K$\alpha$ doublet and the \ion{Fe}{xxv} triplet. To assess this in more detail, the input spectrum of our simulation included two Gaussians for the Fe K$\alpha$, at 1.9399 \AA\ for the Fe K$\alpha_1$ and at 1.9357 \AA\ for Fe K$\alpha_2$, respectively, with a 1:2 ratio \citep{Kaastra_Mewe1993}, and four Gaussians for the \ion{Fe}{xxv}, with line centroids as in \citet{Drake1988} and a flux ratio of 2:1:1:2 ($w$:$x$:$y$:$z$). The width of all the lines was fixed to 0.0007 \AA\ ($\sim$2 eV).

\xrism\ will be provided with the soft X-ray spectrometer Resolve, with a nominal energy resolution of 5--7~eV in the 0.3--12~keV bandpass. We used the ancillary and response files of \textit{Hitomi}/SXS \citep{Kelley2016} for the energy resolution requirement of 7~eV. Simulations show that an exposure of only 300 s (comparable with the pulse period of 293 s) is sufficient to clearly detect the Fe K$\beta$ line with a significance of $\alpha=1.8$, corresponding to 83\% of positive detection probability, with a measured Fe K$\beta/$K$\alpha$ ratio of $0.17^{+0.11}_{-0.09}$. With an exposure of 2.5 ks, the probability of a positive detection of the Fe K$\beta$ line raises
up to $>99.99\%$ ($\alpha=22$). The Fe K$\alpha$ doublet is resolved, while amongst the lines of \ion{Fe}{xxv} only the $f$ line is clearly resolved.

\begin{figure}
   \centering
   \includegraphics[width=\hsize]{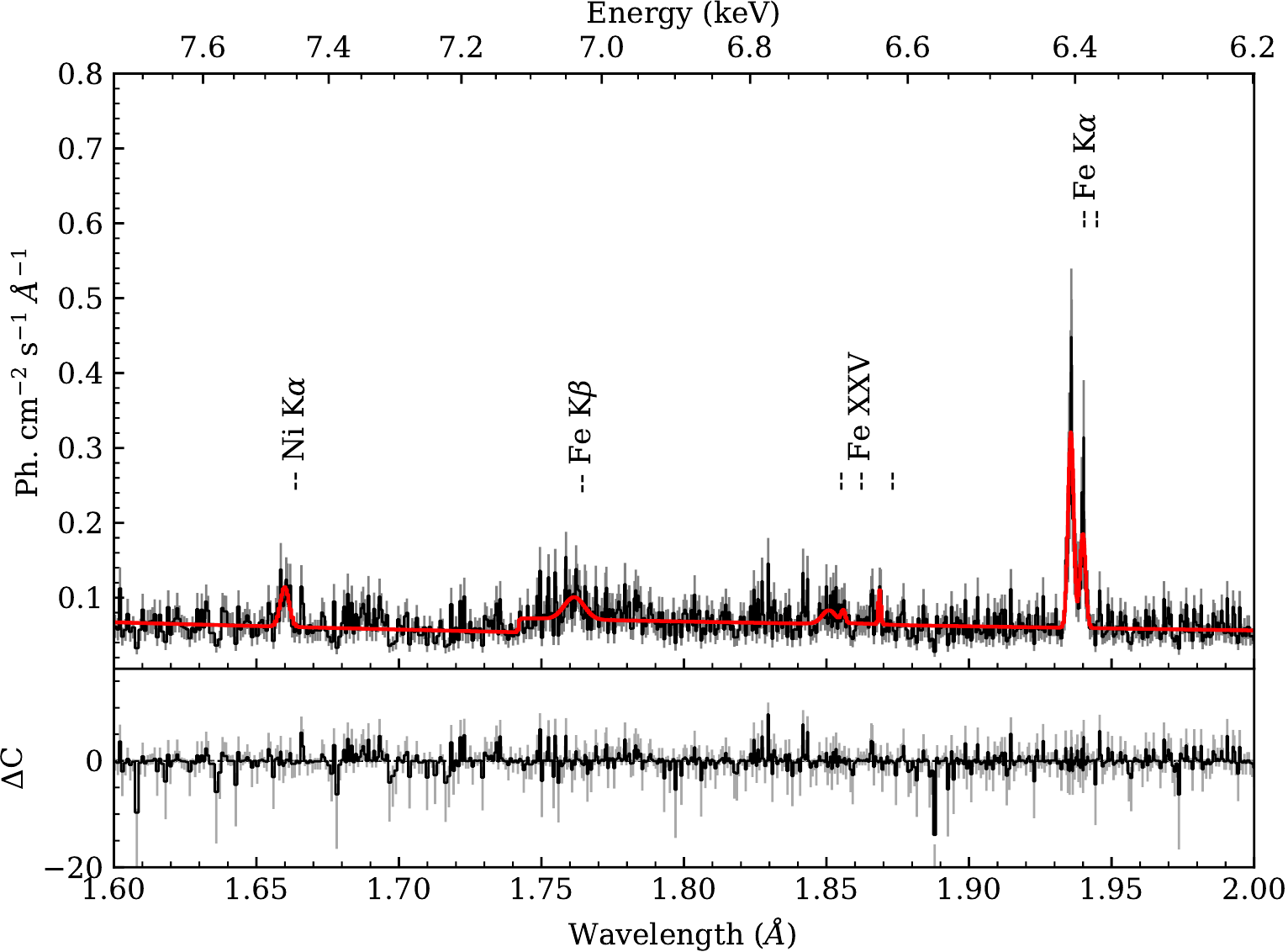}
      \caption{Simulated spectrum of the Fe region with the \athena/X-IFU and best fit model, with residuals in the lower panel. Exposure time of 300 s, data binned with a minimum of 15 counts/bin.}
     \label{fig:athena_sim}
   \end{figure}

\athena will be equipped with the X-ray Integral Field Unit \citep[X-IFU,][]{Barret2018}, a cryogenic X-ray spectrometer working in the energy range 0.2--12 keV, with a nominal energy resolution of 2.5 eV up to 7 keV. Moreover, thanks to the higher collecting area of \athena (1.4 m$^2$ at 1 keV), high quality spectra will be acquired in much shorter exposures. Also for the \athena{}/X-IFU, we performed a 300 s simulation of the Fe region (Fig.\,\ref{fig:athena_sim}). Running the BB algorithm on the simulated spectrum, the K$\beta$ line is detected with $\alpha=9$, corresponding to 99.99\% probability of positive detection.
If the exposure times is increased up to 2.5 ks, then the K$\beta$ line is detected with a significance of $\alpha=69$. 
The measured intensity ratio between the Fe K$\beta$ and Fe K$\alpha$ is $0.16^{+0.10}_{-0.08}$. The Fe K$\alpha$ doublet is fully resolved, as well as the $f$ line of \ion{Fe}{xxv}. The $i$ line, which is made by two lines \citep[$(x+y)$ in the nomenclature of][]{Gabriel1972}, is partially resolved, with the most energetic one blended with the $r$ line. 

\begin{figure}
   \centering
   \includegraphics[width=.9\hsize,height=.7\hsize]{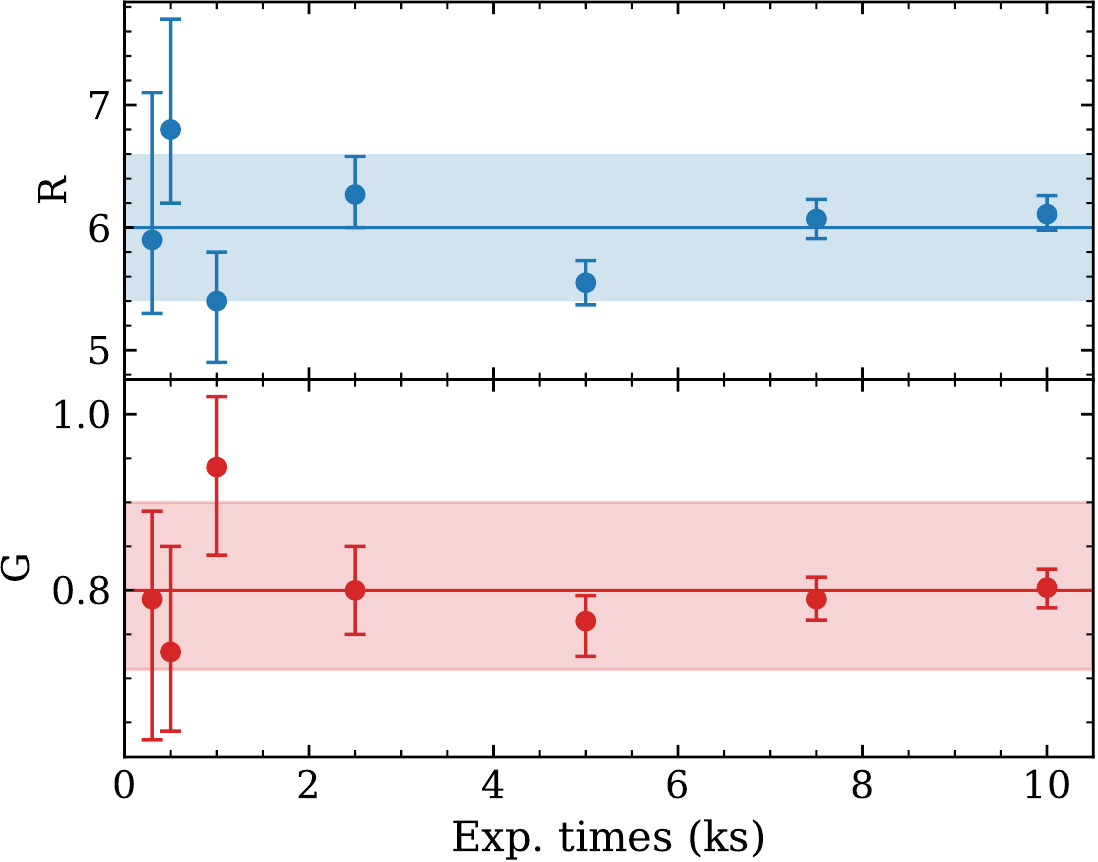}
      \caption{$R$ and $G$ ratios for the He-like triplet of Si as obtained from simulations with \athena/X-IFU with different exposure times. Solid lines correspond to the best fit values with the error ranges given by the coloured areas obtained from the present work.}
     \label{fig:athena_sim_RG}
   \end{figure}

\athena's capabilities will significantly improve also plasma diagnostic, even at shorter exposures. To test how well we can determine $R$ and $G$ ratios, we performed simulations with \athena/X-IFU at different exposure times. Fig.\,\ref{fig:athena_sim_RG} shows the ratios of the Si regions at different exposures, in comparison with the ratios obtained from the analysis of the 45.88 ks \chandra/HETGS observational data set. With an exposure of only 2.5 ks the uncertainties on $R$ and $G$ are reduced of the $\sim$50\%. Longer exposures reduce consistently the errors on $R$ and $G$, from $\sim$10\% up to $2$\% of their absolute values. 

All the discussed simulations with \athena/X-IFU were performed using standard response matrices and background files\footnote{Response matrices for the \athena/X-IFU can be found at: \url{http://x-ifu-resources.irap.omp.eu/PUBLIC/RESPONSES/CC_CONFIGURATION/}. Background files are available at: \url{http://x-ifu-resources.irap.omp.eu/PUBLIC/BACKGROUND/CC_CONFIGURATION/}}. A more thorough exploration of possibilities to observe Vela X-1 with \athena, including a detailed modelling of the effects of defocussing necessary to avoid pile-up for bright X-ray binaries and the right choice of event grades to address certain scientific questions, is beyond the scope of this work and will be addressed in a dedicated publication.

Overall, the achievement of good-quality spectra with such short exposure times imply that the lines can be traced on shorter timescales, i.e., of the same order of magnitude as the pulsar period. Moreover, because of \athena’s resolution, the energy of the Fe K$\alpha$ line can be better constrained so that we can be able to determine the ionisation stage of iron with a higher precision. It is clear, then, that upcoming X-ray satellites will considerably improve the knowledge of HMXBs, of stellar winds and, in general, of any kind of astrophysical plasma, as well remarked by \citet{xrism2020}.

\section{Conclusions}
\label{sec:conclusion}
We conducted, for the first time, X-ray high-resolution spectroscopy of Vela X-1 at the orbital phase $\phi_\mathrm{orb}\approx0.75$, i.e., when the line of sight is going through the photoionisation wake that trails the neutron star along the orbit. 

The data did not show any significant variability of the continuum for the duration of the observation. A blind search for spectral features lead us to detect emission lines from Fe, S, Si, Mg, Ne, and, to a lesser degree, from Al and Na. We detected and identified five narrows RRCs (\ion{Mg}{xi-xii}, \ion{Ne}{ix-x}, \ion{O}{viii}) and He-like triplets of S, Si, Mg and Ne. 

From plasma diagnostic techniques and from fits with photoionisation models from CLOUDY and SPEX, we conclude that the plasma at this orbital phase is mainly photoionised, but data suggest the presence of at least another component, with a smaller ionisation parameter. The presence of a collisional component cannot be excluded, as well as a mixture of ionised and collisional phases. This is in agreement with the idea of colder and denser clumps of matter, embedded in the hot, optically-thin wind of the donor star. The complex geometry of the system is also reflected by the spread of the distribution of the Doppler velocities, as well as in the indetermination of the emission region. 

The future X-ray instruments \athena/X-IFU and \xrism/Resolve will considerably enhance the detection and the resolution of spectral features. We showed through simulations that, thanks to higher energy resolutions, they will resolve single lines in the Fe K$\alpha$ doublets and \ion{Fe}{xxv} triplet and, thanks to higher collecting areas, will allow plasma diagnostic for time scales as short as few hundreds of seconds.

\begin{acknowledgements} 
Authors acknowledge financial contribution from the agreement ASI-INAF n.2017-14-H.0, from INAF mainstream (PI: T. Belloni). VG is supported through the Margarete von Wrangell fellowship by the ESF and the Ministry of Science, Research and the Arts Baden-W\"urttemberg. SB acknowledges financial support from the Italian Space Agency under grant ASI-INAF 2017-14-H.O. Work at LLNL was performed under the auspices of the U.S. Department of Energy under contract No. DE-AC52-07NA27344 and supported through NASA grants to LLNL. 

This research has made use of NASA's Astrophysics Data System Bibliographic Service (ADS) and of ISIS functions (\texttt{isisscripts})\footnote{\url{http://www.sternwarte.uni-erlangen.de/isis/}} provided by ECAP/Remeis observatory and MIT. For the initial data exploration, this research used the \chandra Transmission Grating Data Catalog and Archive\footnote{\url{http://tgcat.mit.edu/}} \citep[\texttt{tgcat};][]{Huenemoerder_2011a}.
This research also has used the following \mintinline{python}{Python} packages: \mintinline{python}{Matplotlib} \citep{matplotlib2007}, \mintinline{python}{Numpy} \citep{numpy2006}, \mintinline{python}{Pandas} \citep{pandas2010}, and the community-developed \mintinline{python}{Astropy} \citep{astropy:2018}. We in particular thank M. Nowak for the implementation of the Bayesian Block algorithm used in this work, M. Guainazzi for his input on \textsl{XRISM} simulations, and I. El Mellah for helpful discussions.
\end{acknowledgements}

\bibliographystyle{aa}
\bibliography{aa_abbrv,mnemonic,references}

\end{document}